\documentclass[aps,prd,reprint,10pt]{revtex4-1}
\usepackage{graphicx} % or graphicx
\usepackage{comment}
\pdfoutput=1
\begin{document}

\title{Maximal CP and Bounds on the Neutron Electric Dipole Moment from \\ P and CP Breaking}
\author{Ravi Kuchimanchi}
\date{March 10, 2012}
\email{raviaravinda@gmail.com}

\begin{abstract}
We find in theories with spontaneous $P$ \itshape and  \normalfont $CP$ violation that symmetries needed to set the tree level strong $CP$ phase to zero can also set all non-zero tree level $CP$ violating phases to the maximal value $\pi / 2$ in the symmetry basis simultaneously explaining the smallness of $\bar{\theta}$ and the largeness of the CKM $CP$ violating phase. In these models we find the one loop lower bound $\bar{\theta} \geq 10^{-11} $ relevant for early discovery of neutron EDM $d_n \geq 10^{-27}$ ecm.  The lower bound relaxes to $\bar{\theta} \geq 10^{-13} $ or $d_n \geq 10^{-29}$ ecm for the case where the  $CP$ phases are non-maximal. Interestingly the spontaneous $CP$ phase appears in the quark sector, not the Higgs sector, and is enabled by a heavy left-right symmetric vectorlike quark family with mass $M$. These results  do not vanish in the decoupling limit of  $M_{H_2^+} > M \rightarrow \infty$ (where $M_{H_2^+}$ is the mass of heavy Higgses at the parity breaking scale) and the age-old expectation that laws of nature (or its Lagrangian) are parity  and matter-antimatter symmetric may be testable  by the above predictions and EDM experiments, even if new physics occurs only at see-saw, GUT or Planck scales.  There is also a region in parameter space with $M_{H_2^+} < M$ where the above  bounds are  dampened by the factor $(M_{H_2^+}/M)^2.$  By using  flavour symmetries and texture arguments we also make predictions for the CKM phase that arises from the maximal phase on diagonalization to the physical basis.  There are no axions predicted in this model.
\end{abstract}

\maketitle
\section{Introduction}
With the discovery of Higgs or Higgs-like boson   the only  standard model parameter that  remains to be determined is the value of the strong $CP$ phase $\bar{\theta}$ that requires the violation of both left-right (or parity P) and matter-antimatter (CP) symmetries. Over the past 6 decades searches for the neutron electric dipole moment (EDM) which is also both $P$ and $CP$ odd  have provided the experimental bound $d_n \leq 1.9 \times 10^{-26}$ ecm at $90 \%$ C.L.~\cite{PhysRevLett.97.131801}. With $d_n \sim 2 \times 10^{-16} \bar{\theta}$~\cite{Pospelov:2005pr} this translates to $\bar{\theta} \leq 10^{-10}$. Several efforts~\cite{Burghoff:2011xk,*Baker:2010, *Beck:2011gw,*Martin:2011,*Serebrov:2009zz}are currently underway to improve the sensitivity of these experiments by two orders of magnitude that can potentially probe $d_n$ down to $10^{-28} ecm$ or $\bar{\theta} \sim 10^{-12}$. Storage ring experiments~\cite{Semertzidis:2011qv} being planned to search for EDMs of the proton and deutron can further this frontier to an equivalent of $d_n \sim 10^{-29} ecm$ or $\bar{\theta} \sim 10^{-13}$.

Since both $P$ and $CP$ are broken in nature we would naively expect $\bar{\theta} \sim \gamma$ where $\gamma$ is the $CP$ violating CKM phase. However experimentally $\bar{\theta} << \gamma \sim 69^o$ which hints at a hidden symmetry, and this large inequality is the well-known strong $CP$ puzzle.

%New physics at the TeV scale such as supersymmetry does not address the strong $CP$ problem  and  it is usually assumed that a separate mechanism such as  $U(1)_{PQ}$ sets $\bar{\theta}$ to zero. Once $\bar{\theta}$ is set to zero $d_n \sim 10^{-32} ecm$ that is radiatively in the standard model alone is too small to be observed in the ongoing and planned neutron EDM experiments.  In order to observe an EDM signal at these experiments we would in effect need one mechanism to turn off the neutron EDM at a high scale ($PQ$ symmetry~\cite{PhysRevLett.38.1440}) and we would additionally need new physics at TeV or so scale to switch it back on. However such beyond the standard model contributions to $d_n$ would in most cases disappear rapidly with the  increasing mass scale $M_{NP}$  of new physics  by the quadratic factor $(v / M_{NP})^2$, where $v$ is the weak scale. For example with supersymmetry as new physics,  $d_n \sim 10^{-25} (TeV/M_{SUSY})^2$ ecm~\cite{Pospelov:2005pr} and becomes smaller than the sensitivity of ongoing experiments  for $M_{SUSY} > (10-100) \ TeV$.

The most popular solution to the strong CP problem is the Peccei-Quinn (PQ) symmetry~\cite{PhysRevLett.38.1440} that dynamically sets $\bar{\theta}$ to zero.  The neutron EDM induced radiatively in the standard model in this case is $d_n \sim 10^{-32} ecm$ and is too small to be observed in the ongoing experiments. There can be beyond the standard model contributions to $d_n$ due to new physics such as supersmmetry but these decrease quadratically as the scale of new physics becomes large.   For example with supersymmetry as new physics,  $d_n \sim 10^{-25} (TeV/M_{SUSY})^2$ ecm~\cite{Pospelov:2005pr} and becomes smaller than the sensitivity of ongoing experiments  for $M_{SUSY} > (10-100) \ TeV$.

%\subsection{Prediction of sizable neutron EDM}
In this work instead of using the $PQ$ symmetry we follow a different line of approach to solve the strong CP problem and show that  the model presented in this paper (along with that in reference~\cite{Kuchimanchi:2010xs}) can lead to a sizable $d_n$ that is discoverable in the ongoing EDM experiments \emph{even if all new physics occurs only at very high scales such as GUT or Planck scales.}

Since violation of both $P$ and $CP$ is fundamental to the existence of a non-zero $d_n$ the approach different from the continuous $PQ$ symmetry is to impose either of these discrete symmetries  to set $\bar{\theta}$ to zero~\cite{Nelson:1983zb,*PhysRevLett.53.329,*FF1984165,*Bento:1991ez,Babu:1989rb,Kuchimanchi:2010xs}. The challenge then is that $P/CP$ must break  without spontaneously generating a strong CP phase at the tree-level so that the experimental constraint $\bar{\theta} << \gamma$ is respected. In the past this has typically required along with vectorlike quarks further  symmetries that generate the Nelson-Barr form of mass matrices as in~\cite{Nelson:1983zb,*PhysRevLett.53.329,*FF1984165,*Bento:1991ez} or mirror parity defined so that it takes known quarks and leptons to 3 additional generations of mirror quark and lepton families as in~\cite{Babu:1989rb}. 

However more recently in~\cite{Kuchimanchi:2010xs} we showed that with no other symmetries imposed and no multiplication of existing families by three generations of mirror families,  just $P$ and $CP$ are sufficient to solve the strong $CP$ problem in the left-right symmetric model with the addition of a full vectorlike quark family. A crucial aspect of this model is that terms and vacuum expectation values (VEVs) that violate P conserve CP and those that violate CP conserve P and therefore the strong CP phase is not generated at the tree level as it is protected by either P or CP for every term.  Our work in~\cite{Kuchimanchi:2010xs}  is the first solution of the strong CP problem in the left-right symmetric model  where $P$ (and not $PQ$ symmetry)is used to set the tree-level $\bar{\theta}$ to zero. (See also \footnote{Potential solutions  without vectorlike quarks were considered in the past for example in reference~\cite{Mohapatra:1978fy}.  However in such models setting the strong $CP$ phase to zero also sets the  $CP$ violating phase to zero in the  CKM matrix with 3 generations and this was overlooked.  This can be seen by noting that  the Jarlskog invariant vanishes.  Thus in order to solve the strong $CP$ problem using parity in the left-right symmetric model  we \itshape need \normalfont to either extend the quark sector such as with vectorlike quarks as we have done or make the theory supersymmetric as  in~\cite{Kuchimanchi:1995rp,*PhysRevLett.76.3490,*Mohapatra:1997su,*Pospelov:1996be}.}.) 

%A crucial step in the proof was to show that $P$ and $CP$ violation can happen in different sectors so that there are no terms that simultaneously violate both. The gauge terms of the high energy Lagrangian conserve both $P$ and $CP$, the fermion mass terms violate $CP$ but not $P$ and the Higgs VEVs violate $P$ but not $CP$.  Thus the $\bar{\theta}$ (and therefore the neutron $d_n$) is not  generated at the tree-level as its protected for each term by either $P$ or $CP$.  

Moreover $\bar{\theta}$ generated radiatively at the one-loop level in this model does not diminish as the scale of $P$ and $CP$ breaking (that is mass scale $M_{H^+_2}$ of  Higgs sector that breaks $P$ and the mass $M$ of vectorlike quarks that are needed to break $CP$)  go to infinity, since CP phases  generated in collusion with Yukawa terms do not respect the decoupling theorem. $d_n$ generated in this model can be much greater than the naive standard model expectation of $10^{-32} ecm$ even if there is no new physics at TeV or 1000's of TeV scales and therefore should be of interest to the ongoing neutron EDM experiments.

%In this work we calculate the lower bound on $\bar{\theta}$  and therefore also on $d_n$ that is generated when $P$ and $CP$ break  so that it can be compared with future experimental results.  

%\subsection{Maximal CP phase} 
We first generalize the model of reference~\cite{Kuchimanchi:2010xs} to allow for spontaneous $CP$ violation (instead of softly through dimension 3 fermion mass terms as in that work) so that we can consider both the case of spontaneous (which could also be more predictive) as well as soft $CP$ breaking.

To our surprise we  find that thus solving  the strong $CP$ problem not only  determines the strong $CP$ phase, but also determines the $CP$ phases in the quark mass matrices   to be maximal  (ie) $\pi / 2$ in the symmetry basis, consistent with the high value of the CKM $CP$ violating phase $\gamma \sim 68^o$ obtained on diagonalizing to the physical mass basis. While several works such as ~\cite{Fritzsch:1995nx, RodriguezJauregui:2001bt,Gronau:1985sp} studying the texture of the quark mass matrices have in the past suggested that $CP$ violation could be maximal (either in the mass matrix or in the Jarlskog invariant), the phase $\pi / 2$ is usually put in by hand and not obtained by symmetries. 

Spontaneous breaking of $CP$ is an attractive idea as it makes $CP$ phases calculable.  In practice however the phases are determined by minimizing the Higgs potential that has several additional parameters due to which we lose predictivity.   What we show is that once we have both P and $CP$ symmetries imposed and solve the strong $CP$ problem, the phase of a CP violating VEV is determined by its P transformation properties independent of the Higgs parameters and  is maximal in the quark mass matrix (that also has vectorlike quarks) in the symmetry basis. See sections~\ref{sec:overview}, \ref{sec:transform},\ref{sec:maximalcp} and~\ref{sec:maximal}.   

It turns out that presence of a  maximal $CP$ phase may be experimentally verifiable by the neutron EDM searches. The lower bound on the strong $CP$ phase  generated  depends on rotations needed to go from the symmetry basis to the physical mass basis. If the $CP$ phase generated in the symmetry basis is $\pi / 2$ (as opposed to an arbitrary number that could be chosen to be the observed CKM phase $\gamma$ for the purposes of calculating a lower bound) then some amount of additional rotation that mixes the real and purely imaginary terms of the mass matrices is needed to obtain the observed $\gamma \sim 68^o$. In a large region of parameter space, this generates a higher lower bound   $\bar{\theta} \geq 10^{-11}$ (or $d_n \geq 10^{-27} ecm$)  and the lower bound would be a couple of orders of magnitude less ($\bar{\theta} \geq 10^{-13}$ or $d_n \geq 10^{-29} ecm$) had the $CP$ phase been arbitrary and not maximal.

The idea of the strong CP problem hinting at a hidden $P$ \itshape and \normalfont $CP$ symmetry that are spontaneously or softly broken as presented in this work (and in~\cite{Kuchimanchi:2010xs})  is verifiable by finding $d_n$ greater than either of the above two lower bounds  for the case where the heavy higgses are heavier than the vector like quark masses (ie) $M_{H^+_2} > M$. While there is a suppression factor of $(M_{H^+_2}/M)^2$ if $M_{H^+_2} < M$ and the lower bounds are then reduced and depend on this factor as well.  See sections~\ref{sec:nedmpi2} and~\ref{sec:nedmany}

%In this model the vectorlike quark masses can be naturally small as shown in~\cite{Kuchimanchi:2010xs} and even in general fermion masses are protected by chiral symmetries.  However scalar masses  can become as heavy as the heaviest mass scales in the theory owing to quadratic divergences.  Thus there are very good reasons to expect $M_{H^+_2} > M$ and even if $M_{H^+_2}$ is smaller than $M$ it may not be much smaller. Thus the prospects for observing $d_n$ in the ongoing and planned EDM searches~\cite{Burghoff:2011xk,*Baker:2010, *Beck:2011gw,*Martin:2011,*Serebrov:2009zz} appear to be reasonably good. 
 
In section~\ref{sec:CKM} we show how texture considerations and flavour symmetries can be used to make predictions for the CKM phase that is generated from the maximal phase on diagonalization from the symmetry basis to the physical basis.

Finally in sections~\ref{sec:comments} and~\ref{sec:conclusions} we present some comments and concluding remarks.

\section{P and CP Properties of $\bar{\theta}=0$ Vacuum \\ Maximal CP Violation}
\label{sec:overview}
%\bfseries \scshape Claim: \normalfont \normalsize If both parity and $CP$ are spontaneously violated in nature, then there is a class of theories where $CP$ violation is generated in the quark mass matrix with the maximal phase of $\pi / 2$  assuming that there is a solution to the strong $CP$ problem 

%\bfseries \scshape Reasoning: \normalfont \normalsize 

We assume $P$ and $CP$ are good  symmetries that are both broken by VEVs of a set of Higgs fields. In general the vacuum will be made of several Higgs fields with VEVs that conserve at the tree level:

\begin{enumerate}
	\item both  $P$ and CP
\item  $CP$ (but not P)
\item $P$  (but not CP)
\item neither $P$ nor CP.
\end{enumerate}

Since $\bar{\theta}$ is both $P$ and $CP$ odd we would expect a strong $CP$ phase to be generated at the tree level by  VEVs that conserve neither $P$ nor CP.    Now if there is a solution to the strong $CP$ problem that sets the tree level strong $CP$ phase to zero,  we would expect that there are no VEVs of the fourth category. That this is in fact  the case   is \itshape proved \normalfont in the next section for a  class of models by examining the minimum of the Higgs potential.  This means in a  class of models, we can visualize the strong $CP$ solving vacuum  to be made of several states, with each state being either $P$ even or $CP$ even or both. This property of the vacuum  can help determine the phase of $CP$ violation that is generated at the tree level as we now show. 

Consider the neutral component $\phi^o$ of a Higgs fields that picks up a VEV $v$. Let us say under $CP$ $\phi^o \rightarrow \phi^{o\star}$ so that any non-real $v$ breaks CP. Likewise we choose the $CP$ properties of quarks and leptons  so that all coupling constants including the Yukawa coupling are real due to CP.

However as discussed, $v$  conserves $P$ if it violates $CP$ so as not to generate a tree-level strong $CP$ phase. Under $P$ if $\phi^o$ transforms non-trivially, then $P$ conservation is a non-trivial  condition that $v$  must satisfy, and its phase gets determined.

Under $P$ we can choose the quarks to transform as $q_{iL} \leftrightarrow q_{iR}$. The Yukawa terms are of the form $\underline{h}_{ij}\bar{q}_{iL} \phi^o q_{R.j} + hc$ with $\underline{h}_{ij}$ real due to CP. Under $P$ if we choose  $\phi^o \rightarrow e^{i\beta} \phi^{o^\star}$  then  $P$ invariance of the Yukawa term together with real $\underline{h}_{ij}$ implies that $e^{i\beta}$ is real. Therefore $\beta = 0$ or $\beta =\pi$ for $\phi^o$ that have Yukawa couplings with the quarks. 

Moreover from the $P$ transformation $\phi^o \rightarrow e^{i\beta} \phi^{o^\star}$  it is easy to see that $\left<\phi^o \right> = |v| e^{i \beta / 2} $ will conserve P, as is needed for VEVs that violate CP.

Thus the phase of the VEV gets determined in terms of the transformation properties of the field under parity. For VEVs that have Yukawa coupling with the quarks $\beta = 0$ or $ \pi$ and so the phase of $\left<\phi^o\right>$ which is $\beta / 2 $ becomes $0$ or $ \pi/2$.  $\beta = 0$ corresponds to the case where both $P$ and $CP$ are conserved by the VEV and $\beta = \pi/2$ violates $CP$ maximally while conserving $P$.  Both types of fields will be present.
 
There are a few points we make before we can say that the $CP$ violation is also maximal in the quark mass matrix in symmetry basis:
\begin{itemize}
\item  The phase generated in the quark mass matrix can be rotated away if there are only the usual 3 families. So there is no CKM phase generated unless the quark content is extended.    This is resolved by having a full vectorlike quark family in the model as proposed in reference~\cite{Kuchimanchi:2010xs}

\item Since there are Higgs fields with both purely imaginary ($\beta = \pi / 2$) and real ($\beta = 0$) VEVs,  if they contribute to the same element of the quark mass matrix then the phase generated in the mass matrix will not be $\pi / 2$ owing to their mixing. We will see in the next section that this does not happen due to the same symmetries (we call these Strong $CP$ Solution Symmetries or S$CP$ for short)  that are imposed to help solve the strong $CP$ problem.  So each term in the mass matrix is purely real or purely imaginary. Thus a maximal $CP$ phase $\pi / 2$ gets generated in the quark mass matrix in the symmetry basis. On diagonalization to the physical basis the real and imaginary parts mix providing a CKM phase consistent with the observations.

\item $CP$ can also be broken spontaneously using a $CP$ odd, $P$ even real singlet.       This case was already discussed  in reference~\cite{Kuchimanchi:2010xs} and its physics is equivalent to $CP$ being violated softly by dimension 3 mass terms (and with no  singlet) as discussed in that work.  The attractive feature of this case (with or without the real singlet) is that no other symmetries (ie SCPs) other than parity and $CP$  need to be introduced to solve the strong $CP$ problem. The flip side  is that it is therefore possible to generate an arbitrary (ie non-maximal) $CP$ phase in the mass matrix.
\end{itemize}
We will estimate lower bounds for $\bar{\theta}$ and the neutron EDM that is radiatively generated in the quark mass matrix  for both the above cases of $CP$ violation....where it is maximal in the quark mass matrix in symmetry basis (see section~\ref{sec:nedmpi2}) as well as where it could take on any value (see~\ref{sec:nedmany}). 

In the next section we set up the $P$ and $CP$ symmetric strong CP solving model and show that $CP$ violation is maximal in the quark mass matrix.  Before we proceed a note on the conventions used which are slightly different from those in this section.  So far we defined $CP$ transformations so that real VEVs conserve $CP$.  However in left-right symmetric theories the convention is to define $P$ and $CP$ such that real VEVs of  doublet $\phi$ conserve $P$ while they could violate $CP$. We revert to this standard convention of left-right symmetric theories in what follows, however since the result of maximal $CP$ violation are convention independent the particular convention chosen does not make a difference to the physics.

\section{Left-Right Matter-Antimatter Symmetric Strong CP  Model \\
$SU(2)_L \times SU(2)_R \times U(1)_{B-L} \times P \times CP \times SCP$}
\label{sec:LRMA}

We begin with the left-right symmetric model~\cite{PhysRevD.10.275,*PhysRevD.11.566,*Senjanovic:1975rk} based on $G_{LR} \equiv SU(3)_C \times SU(2)_L \times SU(2)_R \times U(1)_{B-L} \times P$  with strong $CP$ solving particle content discussed in reference~\cite{Kuchimanchi:2010xs} where $CP$ was imposed on dimension 4 terms and broken softly by dimension 3 mass terms,  and increase the symmetry to include $CP$ as a good symmetry of the Lagrangian. Therefore a second Higgs bidoublet is added so that it can receive a $CP$ breaking VEV. 

The Higgs fields thus consists of the usual $SU(2)_R$ triplet, $SU(2)_L$ singlet $\Delta_{R}$ (and parity related $\Delta_{L}$) and  two bi-doublets $\phi_s$ and $\phi_a$ that transform as $(1,2,2,0)$ under $G_{LR}$. The Higgs bi-doublets and triplets are represented as complex $2 \times 2$ matrices (with the triplet matrices being traceless) in the relevant iso-spaces, as is usual in left-right symmetric model.

The quark content of the model  is the same as in~\cite{Kuchimanchi:2010xs}  where in addition to the usual 3 light chiral families we have a vectorlike $SU(2)_L$ doublet family $Q_{4L}$ and $Q'_R$ that transforms as (3,2,1,1/3), and a parity related vectorlike $SU(2)_L$ singlet ($SU(2)_R$ doublet) family  $Q_{4R}$ and $Q'_L$ that transforms as (3,1,2,1/3). The prime on the quarks is being used in our notation to clearly label the mirror component of the vectorlike family.   

Thus there are two heavy top and two heavy bottom quarks in the model, one pair from the vectorlike doublet and the other from the vectorlike singlet family.  Note that instead of $\Delta_{L,R}$,  $SU(2)_{L,R}$ doublets $\chi_{L,R}$ can also be used to break parity and the results in this work will apply to them them as well. There are the usual 3 generations of leptons and vectorlike lepton families need or need not exist. We do not explicitly mention the lepton sector in this work,  other than  briefly in the second comment of section~\ref{sec:comments}.

\subsection{P and CP transformations}
\label{sec:transform}
Under Parity $SU(2)_L \leftrightarrow SU(2)_R, \ \phi_{a,s} \rightarrow \phi_{a,s}^{\dagger}, \
Q_{iL} \leftrightarrow Q_{iR}, \ Q'_L \leftrightarrow Q'_R, \ \Delta_L \leftrightarrow \Delta_R,
$ Note that with these assignments real VEV's for neutral (i.e. diagonal) components of bidoublets $\phi_{a,s}$ will not contribute to P violation. $Q_{iL,R}$ with $i=1$ to $4$ correspond to the usual 3 generations of quarks and the normal chiral components $Q_{4L,R}$ of the vectorlike quarks. 

We now choose the $CP$ transformation  such that a real VEV for $\phi_a$ will break CP. Under $CP$ we require $\phi_a \rightarrow - \phi_a^{*}, \ \phi_s \rightarrow \phi_s^*, \ Q_{iL,R} \rightarrow CQ^{*}_{iL,R}, \ Q'_{L,R} \rightarrow CQ'^{*}_{L,R}, \ \Delta_{L,R} \rightarrow \Delta^{*}_{L,R}$

With these transformations note that $CP$ invariance of the Lagrangian implies coefficients of all terms (including the Yukawa potential) with an odd (even) number of $\phi_a$ will be purely imaginary (real).   Additionally, the Yukawa matrices must be Hermitian (due to $P$ invariance), dimension 3 quark mass matrix with direct mass terms involving vector-like quarks must be real symmetric (due to $CP$ and P), and dimension 2 Higgs mass parameters must be real (due to P).

%Note that a term such as $\mu_{as}^2 Tr \phi_a^\dagger\phi_s + h.c.$ with $\mu_{as}$ purely imaginary due to $CP$ and purely real due to $P$ will not be allowed if both $P$ and $CP$ invariance is imposed. However a $\mu_{as}^2$ that conserves just $P$ and thus violates $CP$ softly will not adversely affect the strong $CP$ solution as $P$ invariance alone determines it to be real.   

We will now discuss a further symmetry  that needs to  be imposed to solve the strong CP problem and see how this also leads to a maximal $CP$ phase in the quark mass matrix in the symmetry basis.  For this we discuss the Higgs potential and show that symmetries can be broken through real VEVs in the model.

\subsection{$Z_2$ Symmetry, Higgs Potential and\\ P Conserving, CP Violating VEV}
  \label{sec:maximalcp}
If all parameters of the Higgs potential are real then all VEVs that minimize the Higgs potential can be naturally real.   Note that a real VEV for $\phi_a$ will break $CP$ and not $P$ and is of interest to us based on arguments of section~\ref{sec:overview}.  

However there are $P$ and $CP$ invariant quartic terms with an odd number of $\phi_a$ that have purely imaginary couplings and are dangerous for example, 
\begin{equation}
i Tr (\lambda' \phi_a^{\dagger} \phi_s + \lambda'' \tilde{\phi_a}^{\dagger} \phi_s) Tr(\Delta_R^{\dagger} \Delta_R - \Delta_L^{\dagger} \Delta_L )+ h.c.
\label{eq:unwanted}
\end{equation}
with $\lambda'$ and $\lambda''$ real and where $\tilde{\phi}_{a,s} \equiv \tau_2 \phi_{a,s}^{\star} \tau_2$.  We thus need to introduce an additional symmetry under which $\phi_a$  and $\phi_s$ transform differently such as $Z_2$ with $\phi_a$ odd (or anti-symmetric) and $\phi_s$ even (or symmetric) under $Z_2$ to prevent this term.  

Under $Z_2$ we have $\phi_a \rightarrow -\phi_a, \ Q_{1L} \rightarrow -Q_{1L}$ and $Q_{1R} \rightarrow -Q_{1R}$ while all other quark and Higgs fields including $\phi_s$ are invariant under $Z_2$.  Note that one generation of quarks is chosen to be odd under $Z_2$ to permit Yukawa couplings with $\phi_a$ and the remaining generations. $Z_2$  ensures that all terms in the Higgs potential are real since non-real $P$ and $CP$ invariant terms such as~(\ref{eq:unwanted}) vanish due to it. 

Note that any symmetry that sets the purely imaginary quartic couplings to zero will work as well and we designate symmetries such as $Z_2$ as Strong $CP$ solution helping symmetries or SCP symmetries for short.
%and therefore all Higgs VEVs will be naturally real.  

%Note that a Higgs potential with all real couplings will in general have both naturally real as well as complex solutions for the minima in different regions of its parameter space. However the regions with  non-real solutions are phenomenologically not relevant since they would lead to $\bar{\theta} \sim 1$, and hence we are left with the region of parameter space where Higgs VEVs are naturally real. 

We will now discuss the Higgs potential in further detail and show how the symmetry breaking can happen to  give rise to real Higgs VEVs and a  Higgs mass spectrum where all the additional Higgses beyond the standard model Higgs are naturally very heavy. 

The most general Higgs potential invariant under $SU(2)_L \times SU(2)_R \times U(1)_{B-L} \times P \times CP \times Z_2$ is of the form
\begin{equation}
V_{Inv.} = V_{tri}(\Delta_L, \Delta_R) + V_{bi}(\phi_s, \phi_a) + \sum_{z=a, s}{V(\phi_z, \Delta_L, \Delta_R)}  
\label{eq:Vinva}
\end{equation}
where due to $Z_2$ and gauge invariance there are no terms such as in~(\ref{eq:unwanted}) that involve both $\phi_s$ \itshape and \normalfont $\phi_a$ as well as $\Delta_{L,R}$, and $V_{bi}$ contains only quartic terms that involve an even number of both $\phi_s$ \itshape and \normalfont  $\phi_a$. In the above $V(\phi_z,\Delta_L, \Delta_R) +V_{tri}(\Delta_L, \Delta_R)$, with $z = a$ or $s$, is the most general Higgs potential of the minimal left-right symmetric model involving only one bi-doublet $\phi_z$ which is given in references~\cite{PhysRevD.44.837,Duka:1999uc}.  $V, V_{tri}$ and $V_{bi}$ are also functions of the mass parameters and other coupling constants of the model and these indices have been suppressed.  All these parameters are real due to $CP$ and $Z_2$ that have been imposed.

We allow for the Higgs potential to have dimension 2 terms that break $CP$ and $Z_2$ softly such as $-\mu_{1as}^2 Tr (\phi_a^\dagger \phi_s) + h.c.$  All the parameters of soft $CP$ and $Z_2$ breaking terms such as $\mu_{1as}^2$ are real due to $P$.  We thus have for the Higgs potential 
\begin{equation}
V_{Higgs} = V_{Inv.} + V_{soft}
\label{eq:Vhiggs}
\end{equation}
where the detailed form of $V_{Inv}$ and $V_{Soft}$ is given in Appendix~\ref{sec:appendix} and has all real parameters.
 
The neutral component of the Higgs fields can pick up VEVs and these are indicated by
\begin{equation}
\left<\Delta_{L,R}\right> = \left(\begin{array}{cc}
0 & 0 \nonumber \\
\delta_{L,R} & 0 
\end{array}\right), \left< \phi_{a,s}\right> = \left(\begin{array}{cc}
\kappa_{a,s} & 0 \nonumber \\
0 & \kappa'_{a,s}
\end{array}\right)
\label{eq:vevs}
\end{equation}

We break $SU(2)_R \times U(1)_{B-L}$ to $U(1)_Y$ in the usual way with the neutral component of $\Delta_R$ picking a large VEV $v_R \equiv \delta_R$ that sets the scale of parity breaking $M_R$.  Due to $SU(2)_R$ invariance of the potential $v_R$ can be chosen to be real and positive without any loss of generality.  Note in equation~(\ref{eq:A1}) that the coupling term $\rho_3 Tr (\Delta_R^\dagger \Delta_R) Tr (\Delta_L^\dagger \Delta_L)$  with $\rho_3$ positive has its lowest value if $\delta_L = 0$ and therefore this term can ensure that $v_R >> \left| \delta_L \right|$. This is the usual way parity breaks in the left-right model.  
 
%To indicate how symmetry breaking can proceed we note that if  $m_a^2$ in the term $m_a^2 \phi_a^\dagger \phi_a$ of $V(\phi_a,\Delta_L, \Delta_R)$ is sufficiently large and positive then the  
We now look at the bi-doublet VEVs.  Without loss of generality we can use $SU(2)_L$ invariance to choose $\kappa_s$ to be  real and positive.  The remaining VEVs $\kappa'_s, \kappa_a, \kappa'_a$ and $\delta_L$ can in general be complex and their phases are determined by the minimization condition $\partial V_{Higgs} / \partial \phi = 0$ for all the  scalar field components $\phi$ in the theory.  

We now show that for a scalar potential with all real parameters  the minimization condition is satisfied when the VEVs of the fields are all real.

We denote with subscripts $Re$ and  $I$ the real and imaginary components of every field -- for example $\kappa_a = \kappa_{aRe} + i \kappa_{aI}$ etc.  $V_{Higgs}$ can now be written as a sum of  products of real and imaginary parts of the fields. However $V_{Higgs}$ itself is a real-valued function as the Hermitian conjugate of every term is also present in it. Consider a term such as $i \mu^2_{1as} \kappa_{aI} \kappa_{sRe}$ with just one imaginary component of the field that can potentially be obtained from the term $-\mu^2_{1as}Tr(\phi_a^\dagger \phi_s)$. Since $\mu^2_{1as}$ is real, the Hermitian conjugate  $-i \mu^2_{1as} \kappa_{aI} \kappa_{sRe}$ obtained from $-\mu^2_{1as}Tr(\phi_s^\dagger \phi_a)$ provides a canceling contribution.  Hence there can be no terms with an odd number of the imaginary components of the fields in the real-valued scalar potential that has all real parameters.

This implies that 
\begin{equation}
 \left. {\partial V_{Higgs}}\over {\partial \phi_{iI}}\right|_{\phi_{1I} =...=\phi_{nI} =0} =0
\end{equation} 
This is because every term that depends on the imaginary components must have at least two of them and when the partial derivative is taken and evaluated at the point where all the imaginary components are set to zero it vanishes.

 Thus we have shown that real VEVs satisfy the minimization conditions. We now show that there exists a region of parameter space where the scalar potential is at the minimum (and not only at an extremum) for real VEVs.

There are several terms in the Higgs potential that depend on the phases of the fields. We choose the sign of the parameter of every such term so that each of these terms makes a negative contribution to the Higgs potential when all the field values are real and positive. For example we choose mass parameters $\mu_{2s}^2, \mu_{2a}^2, \mu_{1as}^2, \mu^2_{2as}, \mu^2_{2sa} \geq 0$ in~(\ref{eq:beta}) and~(\ref{eq:A4}), and coupling parameters such as $\lambda_{2s}, \lambda_{2a}, \lambda_{4s}, \lambda_{4a}, \lambda_{3as}, \alpha_{2a}, \alpha_{2s}, \rho_4 \leq 0$ in~(\ref{eq:beta}) and~(\ref{eq:A3}), as well as the $\beta$'s $\leq 0$ in~(\ref{eq:beta}).

With this choice of parameter space it is easy to see that the minimum of the Higgs potential will be attained when all the fields $\kappa_{a,s}, \kappa'_{a,s}$ and $\delta_{L,R}$ pick up real positive VEVs (and not any other phases) since each and every term of the potential is independently minimized with this choice.      

Thus we have proved without any fine-tuning that there exists a region of parameter space where the VEVs are real. In reality the parameter space of real VEVs is much bigger -- we need to evaluate the second derivatives of the scalar potential at the extremum point with real VEVs and diagonalize it.  The condition that every mass eigenvalue so obtained is positive (unless it is zero because it is an would-be Goldstone mode) is the only condition that the parameters would need to  satisfy for the extremum to be a minimum.  In fact with only one bi-doublet it is known that a scalar potential with real parameters will only have real VEVs (unless there is a lot of fine-tuning)~\cite{PhysRevD.44.837}. With two bi-doublets as we now have, there maybe some space for non-real VEVs, but a significant region of parameter space will give rise to real VEVs.  The regions with non-real VEVS are phenomenologically ruled out as they would give rise to a tree-level $\bar{\theta}$ and we are left with regions of parameter space with real VEVs.

The remaining minimization conditions $\partial V_{Higgs} / \partial \phi_{iRe} = 0$ have to  be solved simultaneously to obtain the real VEVs in terms of the parameters of the potential. 

Note that the natural value for   mass parameters ($\mu$) of \itshape all \normalfont the dimension 2 terms of the Higgs potential in Appendix~\ref{sec:appendix} is the parity breaking scale $M_R$, which is heavier than the electro-weak scale and can even be as large as the Planck scale.  Thus the natural values for all the VEVs will be either zero or the scale $M_R$. To get a VEV at the electro-weak scale we need to fine-tune parameters in one combination of the minimization equations (say in $\partial V_{Higgs} / \partial \kappa_s = 0$) so that order $M_R$ terms cancel giving rise to  the weak scale.   This is the usual fine-tuning that is needed in the minimal left-right symmetric model or in any non-supersymmetric theory owing to the problem of quadratic divergence.  It ensures an electro-weak scale VEV for $\kappa_s$ and one light Higgs boson which is identified with the standard model Higgs boson. 

If we do not allow for any other fine-tuning the other Higgs bi-doublet $\phi_a$ cannot pick up a VEV in the absence of soft $CP$ and $Z_2$ breaking terms in $V_{soft}$.  %Fine-tuning without introducing $V_{soft)$ can provide a VEV to $\kappa_a$ (just like it does for $\kappa_s$) but this will lead to additional light Higgs bosons and is in any case unnatural.

Therefore we have introduced $V_{soft}$ as a perturbation so that terms such as $\mu^2_{1as} Tr (\phi_s^\dagger \phi_a) + hc$ will induce a VEV to $\phi_a$ from the VEVs of $\phi_s$. Once we substitute for $\delta_R$ in the Higgs potential of Appendix~\ref{sec:appendix} then the terms involving $\kappa_a$ for example can be written in the form 
\begin{equation}
V = -\mu_{1as}^2 \kappa_a \kappa_s + m_{eff}^2 \kappa_a^2
\label{eq:eff}
\end{equation}
where in order to illustrate the process we have set $\mu^2_{2as}, \mu^2_{2sa}$ in $V_{soft}$  to zero.  $V_{bi}$ can be neglected, since for electroweak scale VEVs, $V_{bi}$ is much weaker than $V_{soft}$. We have also neglected terms involving $\kappa'_a$ such as $\kappa_a \kappa'_a$ for purposes of illustration. In the above $m^2_{eff}$ depends on terms such as $\mu^2_{1a}$ in the Higgs potential of Appendix~\ref{sec:appendix} and the VEV $\delta_R$. Further, $\mu_{1as}^2, m_{eff}^2, \kappa_a$ and $\kappa_s$ are all real.

Minimizing the above potential for $\kappa_a$ gives $\kappa_a = (\mu_{1as}^2 / 2 m_{eff}^2) \kappa_s$, where the magnitude of term in the brackets is $< 1$ as it was treated as a perturbation and can be as low as $\sim 10^{-4}$ (since $\kappa_a$ is responsible for Yukawa couplings between first generation of quarks and the rest of the generations).   We can see that even if $\mu_{1as}, m_{eff} \sim M_R$, $\kappa_a$ is naturally small at the electroweak scale set by  $\kappa_s$. 

In general both $\kappa_a$ and $\kappa'_a$ as well as the rest of the terms in $V_{soft}$ will be present. In the potential terms that are quadratic in the fields $\kappa_a$ and $\kappa'_a$ as in equation~(\ref{eq:eff}) will be the only relevant terms in the leading order, and will include cross-terms such as $\kappa_a \kappa'_a$.   Solving minimization equations simulatenously these VEVs can be obtained in terms of liner combinations of the first bi-doublet VEVs $\kappa_s$ and $\kappa'_s$ without any fine-tuning.

The case with one bi-doublet is considered in detail in reference~\cite{PhysRevD.44.837}.  We have shown how the second bi-doublet can naturally pick up VEVs from the first bi-doublet VEVs.

Since there is no fine-tuning other than what is usual for the standard model Higgs mass, the natural scale of all the other Higgs boson masses is $\sim M_R$. 

$\delta_L$ picks up a VEV of the order $v^2 / v_R$ where $v$ is the weak scale, owing to the $\beta$ terms in the potential in equation~(\ref{eq:beta}) in the Appendix~\ref{sec:appendix} as is usual in the left-right symmetric model. The VEV induced for $\delta_L$ is also naturally real as the parameters of the potential are real.

In the next section we look at the fermion mass matrices which are obtained from the Yukawa matrices and real VEVs. 

In place of $V_{soft}$ we can also use a complex singlet to spontaneously break CP without need for any dimension 2 soft CP breaking terms.  This is discussed in section~\ref{sec:domain}.  In the next section we show how real VEVs lead to the solution of the strong $CP$ problem as well as to a maximal $CP$ Phase.    

\subsection{No Strong CP Phase, Maximal CP Phase!}
\label{sec:maximal}
Since Yukawa matrices are Hermitian due to parity, real Higgs VEVs imply that quark mass matrices $M_u$ and $M_d$ are also Hermitian and $\bar{\theta} = Arg Det M_u M_d$ vanishes and the strong $CP$ problem is solved at the tree level. This can be explicitly seen below from the Hermitian form of the mass matrices in equation~(\ref{eq:massmatrixZ2}) obtained using~(\ref{eq:Z2Yukawa}) and~(\ref{eq:massterms})

To note how the maximal CP phase arises recall that under $CP$ $\phi_a \rightarrow -\phi_a^\star$, while the rest of the fields transform in the usual way as given in section~\ref{sec:transform}.  $CP$ conservation implies that the Yukawa couplings involving $\phi_a$  have to be  purely imaginary while the rest of the Yukawa couplings are real. Since the VEVs are real this implies that a purely imaginary contribution is made to the quark mass matrix by Yukawas involving $\phi_a$.

Moreover, the same quark pairs coupling to $\phi_a$ cannot also couple to $\phi_s$ since  $\phi_a$  and $\phi_s$ transform differently under $Z_2$ (or in general under any  SCP).  That is if  the purely imaginary Yukawa coupling $i \bar{Q}_{iL} \phi_a Q_{jR}$ is permitted by $Z_2$, for the same $i, j$ we cannot also have the real Yukawa term $\bar{Q}_{iL} \phi_s Q_{jR}$. This implies that after symmetry breaking through real VEVs for $\phi_a$ and $\phi_s$, the quark mass matrix will not only be Hermitian but will also have entries that are either purely real or purely imaginary.  Thus in the symmetry basis we see that $CP$ violation is maximal -- ie the phase is $\pi/2$. It is interesting that the vanishing of tree-level $\bar{\theta}$ (which motivated the $Z_2$ symmetry) implies a maximal phase for $CP$ violation in the quark mass matrix in the symmetry basis.  

Concretely if the first generation quarks are odd under $Z_2$ and the rest are even, the most general Yukawa terms invariant under $P \times CP \times Z_2$ take the form:
\begin{eqnarray}
\sum_{j=2-4}&i[ \bar{Q}_{1L} (\underline{h}_{1j} \phi_a + \underline{\tilde{h}}_{1j}\tilde{\phi_a}) {Q}_{jR} - \bar{Q}_{jL} (\underline{h}_{1j} \phi_a + \underline{\tilde{h}}_{1j}\tilde{\phi_a}) {Q}_{1R}]   \nonumber \\  +&\bar{Q}_{1L} (\underline{h}_{11}\phi_s + \underline{\tilde{h}}_{11} \tilde{\phi}_s) {Q}_{1R} + \bar{Q}'_L (\underline{h}'\phi_s^{\dagger} + \underline{\tilde{h}}' \tilde{\phi}_s^{\dagger}) Q'_R \nonumber \\ + & \sum_{i,j=2-4}	 \bar{Q}_{iL} (\underline{h}_{ij} \phi_s + \underline{\tilde{h}}_{ij}\tilde{\phi_s}) {Q}_{jR}  + h.c. 
\label{eq:Z2Yukawa}
\end{eqnarray}
where in the last term $\underline{h}_{ij} = \underline{h}_{ji}^\star, \ \underline{\tilde{h}}_{ij} = \underline{\tilde{h}}_{ji}^\star$ due to $P$, and  $\underline{h}_{ij},\underline{\tilde{h}}_{ij} $ are real  due to CP for any $i, j$ in all terms.  The underline on the Yukawa terms like $\underline{h}_{ij}$ has been used to signify that these Yukawas are in the symmetry basis. 

Due to vectorlike quarks there are also the direct quark mass terms 
\begin{equation}
M_i \ \bar{Q}_{iL}  {Q'}_{R} \ +  \ M^\star_i \bar{{Q'}}_{L}  Q_{iR} \ + \ h.c
\label{eq:massterms}
\end{equation}

where the form is due to parity and $M_i =M^\star_i$ are real due to $CP$ invariance. As we can see in~(\ref{eq:massmatrixZ2}) below $M_1 \neq 0$ is needed to ensure that the $CP$ phase generated is not trivially rotated away. $M_1$ breaks $Z_2$ softly.

Using~(\ref{eq:Z2Yukawa}) and~(\ref{eq:massterms}) after electro weak symmetry breaking the up sector quark mass terms are given by
\begin{math}
\left({{\bar{u}}_L, \bar{u}'_{L}}\right) \ M_u \ 
\left(\begin{array}{c}
{u}_R  \\ 
u'_{R}  \ 
\end{array}\right)
\end{math}
with
\begin{equation}
M_u =
\left(\begin{array}{cccc|l}      
\underline{h}^u_{11} v  & i\underline{h}^u_{12} v & i\underline{h}^u_{13} v & i\underline{h}^u_{14} v &  M_1 \\       
-i\underline{h}^u_{12} v  & \underline{h}^u_{22} v  & \underline{h}^u_{23} v & \underline{h}^u_{24} v&  M_2 \\
-i\underline{h}^u_{13} v & \underline{h}^u_{23} v  & \underline{h}^u_{33} v & \underline{h}^u_{34} v& M_{3} \\ 
-i\underline{h}^u_{14} v & \underline{h}^u_{24} v  & \underline{h}^u_{34} v & \underline{h}^u_{44} v& M_{4} \\ 
\hline 
M_1 & M_2 & M_{3} & M_{4} &  \underline{h}'^u v 
\end{array}\right)
\label{eq:massmatrixZ2}
\end{equation}
 where $\bar{u}_L$ is the $1 \times 4$ row vector $\left(\bar{u}_{1L},..., \bar{u}_{4L}\right)$ and from ~(\ref{eq:Z2Yukawa}), $\underline{h}^u_{1j} v = \underline{h}_{1j} \kappa_a + \tilde{\underline{h}}_{1j} \kappa'_a$ for $j \neq 1, \  \underline{h}^u_{ij} v = \underline{h}_{ij} \kappa_s + \tilde{\underline{h}}_{ij} \kappa'_s$ for $i$ and $j \neq 1$ as well as for $i = j = 1$, and $\underline{h}'^u v = \underline{h}' \kappa_s + \underline{\tilde{h}}' \kappa'_s$.  Note that $\{\kappa_{a,s}, \kappa'_{a,s}\}$ are the diagonal elements of $\left<\phi_{a,s}\right>$ and are real, and $v^2 = \kappa^2_a + \kappa'^2_a +\kappa^2_s + \kappa'^2_s$.  Note that we are using a convention where $v \sim 173 GeV$  so that a $\sqrt{2}$ has been absorbed in its definition.
 
   The down quark mass matrix $M_d$ is similar to the up quark mass matrix with  $u \rightarrow d$ and $\kappa_{a,s} \leftrightarrow \kappa'_{a,s}$ in equations determining them.  We use the notation $\underline{H}_u$ for the Hermitian upper $4 \times 4$ sub-matrix of $M_u$ in~(\ref{eq:massmatrixZ2}) divided by the VEV $v$.  Likewise $\underline{H}_d$ is the upper $4 \times 4$ submatrix of $M_d$ divided by $v$.  The underline once again indicates that these matrices are in the symmetry basis.

Note that the above matrix has three light quark mass eigenvalues at the weak scale corresponding to the usual 3 light chiral families and 2 heavy quark mass eigen values $\sim \pm \sqrt{\Sigma |M_i|^2}$ (up to electro-weak corrections) -- thus there are two heavy  down quarks one each from the vectorlike doublet and the vector like singlet family.  Similar is the case with the up sector.

\vspace{1in}
\subsection{Role Played by $P, CP,$ and $Z_2$}
To summarize, 
\begin{itemize}
\item $P$ implies that the Yukawa matrices are Hermitian.  
\item CP implies that the Yukawa parameters are either purely imaginary (if they couple to $\phi_a$) or purely real (if they couple to $\phi_s$).  
\item In addition to $CP$, $Z_2$ is needed to ensure that there are no non-real parameters in the scalar potential. Parameters of dimension 2 soft $CP$ and $Z_2$ breaking terms are real due to P. Since all parameters of the Higgs potential are real all VEVs can be naturally real.
\item The quark mass matrices that depend on the product of Hermitian Yukawas with real VEVs are also Hermitian thus ensuring that $\arg \det M_u M_d =0$ and solving the strong CP problem.
\item Real VEVs multiplying the purely imaginary or real Yukawa terms  give rise to purely imaginary or real quark mass matrix terms, simultaneously ensuring a maximal CP phase in the quark mass matrices in the symmetry basis. Here $Z_2$ ensures that the same quark pairs do not couple to both $\phi_a$ and $\phi_s$, keeping the purely imaginary and real terms separate.
\item The vectorlike quarks ensure that the $CP$ phase generated cannot be trivially rotated away.
\item Real VEVs conserve either $P$ or $CP$ or both. For example real VEVS of $\Delta_{L,R}$ conserve $CP$ while breaking $P$,  real VEVs of $\phi_a$ conserve $P$ while breaking $CP$, and real VEVs of $\phi_s$ conserve both $P$ and $CP$.
\end{itemize}

\subsection{Spontaneous CP Phase in Fermion Sector}
\label{sec:spontfermion}
An interesting feature of equation~(\ref{eq:Z2Yukawa})  is that we can redefine the first generation quark fields $Q_{1L,R} \rightarrow -i Q_{1L,R}$ so that the phase $i$ gets absorbed so that it disappears from the Yukawa terms in~(\ref{eq:Z2Yukawa})  and appears in terms of the redefined fields in the dimension 3 mass term as $i M_1 (\bar{Q}_{1L} Q'_R - \bar{Q}'_L Q_{1R})$. 

Also since all terms of the Higgs potential are real and the Higgs VEVs are also real, the model has a complex phase  only in the dimension 3 fermion mass term and at this stage is almost exactly like the model of~\cite{Kuchimanchi:2010xs} where $CP$ is broken softly by fermion mass terms.  The only difference is that it is more predictive -- $CP$ phase has been determined to be $\pi / 2$ and $CP$ violation is actually spontaneous since there is a way of defining $CP$ so that the Lagrangian is $CP$ symmetric, as that's how we constructed the model.  However at the tree-level the spontaneous phase is generated in the quark mass matrix and not in the Higgs terms.  

That spontaneous $CP$ violation in multi-Higgs doublet models can appear at the tree level in the fermion sector instead of the Higgs sector was noticed in reference~\cite{Ferreira:2010bm} in a more academic context and its relevance to physical things like the strong $CP$ problem and a maximal $CP$ phase was not considered.

\subsection{Domain Wall Problem}
\label{sec:domain}
Spontaneous $CP$ violation at the weak scale (due to VEV of $\phi_a$)  can lead to the domain wall problem~\cite{Zeldovich:1974uw}. We have addressed this automatically since we have dimension 2 soft CP breaking terms $V_{soft}$. Moreover, the scale of the mass parameters in $V_{soft}$ is naturally of the order $M_R$.

However if we do not want any soft $CP$ breaking,we can add complex singlet fields $\sigma_a$ that can pick up large $CP$ violating real VEVs that conserve P, such that under CP $\sigma_a \rightarrow - \sigma_a^\star$, under $P$ $\sigma_a \rightarrow  \sigma^\star_a$ and under $Z_2$ $\sigma_a \rightarrow -\sigma_a$. As before $Z_2$ ensures that dangerous $P$ and $CP$ symmetric terms of the type $i \sigma_a(\Delta^\dagger_R\Delta_R - \Delta^\dagger_L\Delta_L) + h.c.$ are absent.  Further if we change the $CP$ transformation of $\phi_a$ to be $\phi_a \rightarrow \phi^\star_a$ then the maximal CP phase is generated due to the real VEV of  $\sigma_a$ from the Yukawa term $i h_a \sigma_a (\bar{Q}_{1L} Q'_R -\bar{Q}'_L Q_{iR}) + h.c.$.  The terms such as $\mu^2_{1as} Tr \phi_a^\dagger\phi_s + h.c.$ with $\mu^2_{1as}$ real, now conserve $CP$ while breaking $Z_2$ softly.  Thus with such a singlet, $CP$ can be violated only spontaneously while $Z_2$ can be broken by dimension 2 terms and conserved by dimension 3 and 4 terms. As before $Z_2$ ensures that the real and purely imaginary contributions do not mix in the symmetry basis in the quarks mass matrix and the $CP$ violation is maximal.   If $\sigma_a$ picks a large VEV higher than the scale of inflation then the domain walls will be washed away resolving the problem.  

The case of a real singlet has already been considered in~\cite{Kuchimanchi:2010xs} and it does not require a $Z_2$ symmetry to be imposed as the above dangerous term is automatically absent if $\sigma_a$ is real and the strong CP problem is solved just by $P$ and $CP$ symmetries without needing additional help. 
 
In what follows we allow real $\mu^2_{1as} \neq 0$ and $CP$ is either broken softly without a singlet, or equivalently there can be the singlet and $CP$ is broken spontaneously without any soft CP violation.  The discussion in the rest of the paper does not depend on the details of this choice.    

\section{Radiative Corrections and Neutron EDM} 
\label{sec:nedm}

We follow the procedure in reference~\cite{Kuchimanchi:2010xs} where  radiative corrections to the quark mass matrix and consequently to $\bar{\theta} = Arg Det M_u M_d$ have been evaluated at the one loop level. 

Since $P$ and $CP$ violating terms will interact with one another in loops we expect $\bar{\theta}$ to be generated radiatively. Without loss of generality we can choose a flavour basis via a common $4 \times 4$ orthogonal transformation $O$  on the up and down, such that $O^{T}_{ij}M_j = \delta_{i4} M$  where $\delta_{i4} = 0$ for $i < 4$ and $\delta_{44}=1$.  In the transformed  basis $H_u$ and $H_d$ are hermitian matrices with complex phases as they are obtained from Hermitian matrices in symmetry basis  by  $H_{u,d} = O^T \underline{H}_{u,d} O$. Correspondingly the mass matrices in~(\ref{eq:massmatrixZ2}) transform as  $M_{u,d} \rightarrow \left(\begin{array}{cc} O^T & 0 \\ 0 & 1 \end{array}\right) M_{u,d} \left(\begin{array}{cc} O & 0 \\ 0 & 1 \end{array}\right)$. We can now apply a common unitary rotation in the light $3 \times 3$ sector to diagonalize the $3 \times 3$ subspace of $H_u$ and rotate away all phases in the $4^{th}$ row (and column) so that it is real symmetric. In this basis $H_d$ is a $4 \times 4$ hermitian matrix with complex coefficients.    Writing explicitly the matrix elements $h^{u,d}_{ij}$ of $H_{u,d}$,  the up and down mass matrices look as follows in this physical basis:
\noindent
\begin{equation}
M_u =
\left(\begin{array}{cccc|l}      
 h^u_{11} v &  0  & 0   & h^u_{14} v&  0 \\       
 0 & h^u_{22} v & 0 & h^u_{24}v  & 0 \\
 0& 0  & h^u_{33} v &  h^u_{34} v & 0 \\
h^u_{14} v & h^u_{24}v  &  h^u_{34} v & h^{u}_{44} v &  M\\
\hline
0 & 0 & 0 & M &  h'_u v 
\end{array}\right)
\label{eq:Md_diag_basis}
\end{equation}
\smallskip
\newline
\noindent
\begin{equation}
M_d =
\left(\begin{array}{cccc|l}      
h^d_{11}v  & h^d_{12}v  & h^d_{13}v   & h^d_{14} v &  0 \\       
h^{d\star}_{12}v  & h^{d}_{22}v  & h^d_{23}v & h^d_{24}v & 0 \\
h^{d\star}_{13}v & h^{d\star}_{23}v   & h^d_{33} v  & h^d_{34}v  & 0 \\
h^{d\star}_{14}v & h^{d\star}_{24}v  & h^{d\star}_{34}v  & h^d_{44}v  &  M\\
\hline
0 & 0 & 0 & M &  h'_d v 
\end{array}\right)
\label{eq:Mu_Md_diag_basis}
\end{equation}
\newline
where in this basis $M_u$ is real symmetric and $M_d$ is Hermitian with complex coefficients $h^{d}_{ji} = h^{d\star}_{ij}$ 

We denote by $\delta M_{u_{ij}}$ the radiative corrections to the $i^{th}$ row and $j^{th}$ column of $M_u$ and evaluate the determinant to the lowest order in $\delta M_u$ and find as in reference~\cite{Kuchimanchi:2010xs} the most significant contribution to
\begin{eqnarray}
\arg \det  (M_u + \delta M_u) =   Im  \sum_{i=u,c,t} {\frac{\delta M_{u_{ii}}}{h^u_{ii}v}}   
\label{eq:det}
\end{eqnarray}

$\delta M_{u_{ii}}$ to one loop was evaluated in reference~\cite{Kuchimanchi:2010xs} and we outline the steps here leading to the same results.  We keep the discussion more general so that the same results also apply to the case of more than one bi-doublet. Note that in reference~\cite{Kuchimanchi:2010xs} we were working in a basis where the $3 \times 3$ sector of $M_d$ was diagonal and so we actually evaluated the contributions to $\delta M_{d_{ii}}$ (instead of $\delta M_{u_{ii}}$) in that work.  However the procedure is the same when we work in the basis where $3 \times 3$ sector of $M_u$ is diagonal.

To generate an imaginary part to $\delta M_{u_{ii}}$  we need to look for Feynman diagrams where there are both complex phases and parity violation.  

\begin{figure}[ht]  
\includegraphics[width=55mm]{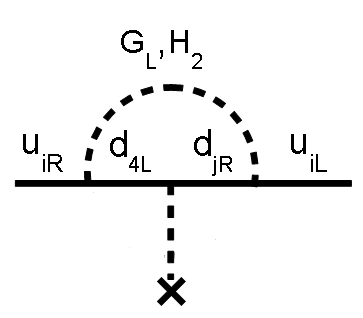}
\caption{Divergent part of radiative contributions from charged would be Goldstone mode and second Higgs  doublet that are related by parity cancels}
\label{oneloophiggscharged}
\end{figure}

In the Feynman gauge the charged would be goldstone boson $G^+_L$ corresponding to $SU(2)_L$ breaking makes a logarithmically divergent contribution to $\bar{\theta}$ owing to the process in Figure~\ref{oneloophiggscharged}. Since it is the  goldstone mode of $SU(2)_L$ transformation its Yukawa couplings are~\cite{Duka:1999uc}  
\begin{equation}
\mathcal{L}_Y = \bar{u}_{iR}h^{u}_{ij} d_{jL} G^+_L -\bar{u}_{iL}h^{d}_{ij} d_{jR} G^+_L + h.c. 
\label{eq:Goldstone}
\end{equation}
However, since among the Higgses only $\phi_{a,s}$ couple to the quarks, and they pick up $P$ symmetric real VEVs, and the Yukawa matrices are Hermitian, there cannot be a net one-loop contribution to $\bar{\theta}$ from the Higgses in the limit that the bidoublets do not couple  to $\Delta_R$ -- whose VEV is the only source of parity violation. 

Therefore the contribution of $G^+_L$ to $\bar{\theta}$ will be exactly canceled by another massless mode in the Higgs bidoublets which we can call $H^+_2$, and it corresponds to a global $SU(2)^{bi}_{R}$ rotation $U^{bi}_{R}$ of the bi-doublets and fermions alone. That is, under this unitary transformation, $\phi_{a,s} \rightarrow \phi_{a,s}U^{bi\dagger}_{R}, \ Q_{iR} \rightarrow U^{bi}_R Q_{iR}$, and $\Delta_R \rightarrow \Delta_R$.   In the symmetry limit $\Delta_R$ Higgses do not couple to the bidoublets other than through $Tr(\phi^\dagger_{s} \phi_{s})Tr \Delta^\dagger_R \Delta_R$, which due to the trace does not convey $SU(2)_R$ breaking to the bi-doublets.

Now if we turn on the $P,CP$ and $Z_2$ invariant coupling terms between the bidoublet and $\Delta_{L,R}$ such as $\alpha_{3s} Tr (\phi^\dagger_{s} \phi_{s} \Delta^\dagger_R \Delta_R + \phi_{s} \phi^\dagger_{s}  \Delta^\dagger_L \Delta_L) + h.c.$ of equation~(\ref{eq:beta}) so that $U^{bi}_R$ is explicitly broken, then when $\Delta_R$ picks up a large VEV, $H^+_2$ will acquire a large mass  $M_{H^+_2} \sim \sqrt{\alpha_{3s}} M_R$, where $M_R$ is the parity breaking right-handed scale. The cancellation of the divergent part of the contribution from $G^+_L$ by $H^+_2$ is unaffected by the mass picked up by $H^+_2$ and a finite contribution to  $\delta M_{u_{ii}}$ is generated due to figure~\ref{oneloophiggscharged} which depends on the product $(H_d H_d H_u)_{ii}$ so that $\bar{\theta}$ due to~(\ref{eq:det}) becomes

\begin{equation}
\bar{\theta}|_{\delta {M_u}} \sim {\frac{1}{16 \pi^2}} \sum_{\stackrel{i =1 \ to \ 3}{j = 1 \ to \ 4}}   \ Im \left(\frac{{h^{d}_{ij}}{h^{d}_{j4}}h^{u}_{4i}}{h^u_{ii}}\right) \ln(M_{H^+_2}/M)
\label{eq:theta_higgs_lowM}
\end{equation}
where $M < M_{H^+_2}$ and the logarithmic dependence is because the divergence of figure~\ref{oneloophiggscharged} beyond $M$  has been canceled beyond scale $M_{H^+_2}$. 

As in reference~\cite{{Kuchimanchi:2010xs}} we also note that if $M > M_{H^+_2}$, then since below the scale $M$ we have the left-right symmetric model, parity protects $\bar{\theta}$ from being generated so that the logarithm above is replaced by the factor $(M_{H^+_2}/M)^2$ which vanishes in the limit the parity breaking scale  $M_{H^+_2} \rightarrow 0$.

Note that the bounds on $\bar{\theta}$ and $d_n$ we are calculating are on their magnitude and so the sign of the quantities is not relevant for these purposes and we are not keeping track of it. Also the bounds are an order of magnitude estimate and not exact.  For example there can be a quark mass dependent coefficient  of order 1 that multiplies the terms within the summation sign of~(\ref{eq:theta_higgs_lowM}).  That is, the coefficient for $j=4$ in~(\ref{eq:theta_higgs_lowM})when the heavy quark is in the inner loop of figure~\ref{oneloophiggscharged} can be different from those for $j = 1 - 3$ for light quarks whose masses can be taken to be zero.  However since these coefficients are both order 1 we have combined the two terms instead of writing them separately. This simplification does not affect our estimates of the lower bounds.  

Keeping these in mind we write in a compact form

\begin{equation}
\bar{\theta}|_{\delta {M_u}} \sim {\frac{1}{16 \pi^2}} \sum_{\stackrel{i =1 \ to \ 3}{j = 1 \ to \ 4}}   \ Im \left(\frac{{h^{d}_{ij}}{h^{d}_{j4}}h^{u}_{4i}}{h^u_{ii}}\right) \ell
\label{eq:theta_F}
\end{equation}
where the factor
\begin{equation}
  \ell = \left\{
  \begin{array}{l l}
    \ln(M_{H^+_2}/M) \geq 1 & \quad for \ M < M_{H^+_2} \nonumber\\
   (M_{H^+_2}/M)^2  & \quad for \ M > M_{H^+_2} \\
  \end{array} \right.
  \label{eq:F}
\end{equation}
The choice of the symbol $\ell$ is to remind us that its a \itshape logarithmic  \normalfont factor for $M < M_{H^+_2}$ and for purposes of calculating lower bounds in this region of parameter space can be chosen to be $O(1)$. We also note that~(\ref{eq:theta_F}) is in the basis where $3\times 3$ submatrix of $M_u$ is diagonal. It can also be written in a basis independent manner as shown in reference~\cite{Kuchimanchi:2010xs} as
 \begin{equation}
\bar{\theta}|_{\delta {M_u}} \sim {\frac{1}{16 \pi^2}} \sum_{\stackrel{i,l =1 \ to \ 3}{j,k = 1 \ to \ 4}}   \ Im \left(h^{d}_{ij}h^{d}_{jk}h^{u}_{kl}(h^{u^{-1}}_{3 \times 3})_{li}\right) \ell
\label{eq:theta_F_inv}
\end{equation}
where  $h^{u^{-1}}_{3 \times 3}$ is the inverse of the $3 \times 3$ submatrix of $H_u$.   The contribution to $\bar{\theta}$ from radiative corrections to $M_d$ can be got from~(\ref{eq:theta_F_inv}) with $u \leftrightarrow d$.
 
Note that the Yukawa terms $L^T_R\tau_2 \Delta_R L_R$  that give Majorana mass to the right handed neutrinos break $SU(2)^{bi}_R$ and  $\alpha_3$   cannot naturally be kept small. Therefore we expect $M_{H^+_2} \sim M_R$.

\subsection{Neutron EDM With Maximal CP Violation}
\label{sec:nedmpi2}
We will now use the formulas in~(\ref{eq:theta_F}) and~(\ref{eq:theta_F_inv})  to evaluate the lower bound for $\bar{\theta}$ in terms of the CKM parameters.  As each element of $M_u$ and $M_d$ in the symmetry basis of~(\ref{eq:massmatrixZ2}) can potentially contribute to $\bar{\theta}$ we can get the lower bound by setting maximum number of these matrix elements to zero while still having sufficient non-zero off-diagonal elements to obtain  the correct CKM mixing angles.

Since we need non-trivial participation from vector-like quarks for $CP$ violation we need to set $M_1, M_4 \neq 0$. In addition we need 3 off diagonal Yukawa couplings $\underline{h}^{u,d}_{ij}$ to obtain the 3 CKM angles if we set $M_2 = M_3 = 0$. At least one these Yukawa terms (but not all) must be from the fourth row or column, or else the the $CP$ phase can be rotated away.

At this stage once we rotate the heavy quarks away and go into the physical basis of equations~(\ref{eq:Md_diag_basis}) and~(\ref{eq:Mu_Md_diag_basis}), the light $3 \times 3$ quark sector will have 3 off diagonal elements (they could be anywhere in up and down  sectors) each coming from a different non-zero Yukawa of the symmetry basis.  So a maximal $CP$ phase $\pi / 2$ is transferred to the light $3 \times 3$ sector and the CKM angle generated is too high at $\gamma \sim \pi/2$.  To get the CKM phase of $68.8^o$, we need one more non-zero Yukawa, so finally we need a texture with 4 non-zero off-diagonal Yukawa terms in the symmetry basis.

For all such textures we can evaluate $\bar{\theta}$ and we find the lower bound $\bar{\theta} \geq 10^{-11} \ell$.   For example, let us take all off-diagonal Yukawas as well as $\underline{h}^u_{44}$ in $M_u$ of(\ref{eq:massmatrixZ2}) to be zero. Let us also take $\underline{h}^d_{44} = \underline{h}^d_{14} = \underline{h}^d_{24} = 0$ so that the only non-zero off-diagonal Yukawas in $M_d$ are $\underline{h}^d_{12}, \underline{h}^d_{13}, \underline{h}^d_{23}$, and  $\underline{h}^d_{34}$. Also we set $M_1, M_4 \neq 0$ and $M_3 = M_4 = 0$.  

We now go into the physical basis by an orthogonal transformation in the $1-4$ plane with $s_1 = sin \theta_1 = M_1 / M$, $c_1 = cos \theta_1$ and $M= \sqrt{M_1^2 + M_4^2}$ we obtain
\begin{equation}
H_u =
\left(\begin{array}{cccc}      
c^2_1 \underline{h}^u_{11} & 0  & 0   & - s_1 \underline{h}^u_{11}  \\       
0  & \underline{h}^u_{22}  & 0 & 0 \\
0 & 0   & h^u_{33}  & 0 \\
- s_1 \underline{h}^u_{11} & 0 & 0  & s^2_1 \underline{h}^u_{11}  \\
\end{array}\right)
\label{eq:Mu_texture}
\end{equation}
\newline
\noindent
\begin{equation}
H_d = 
\left(\begin{array}{cccc}      
0 \ & i c \underline{h}^d_{12}   & ic_1 \underline{h}^d_{13}+ s_1 \underline{h}^d_{34}  & 0 \\    
\star \ & \underline{h}^d_{22}  & \underline{h}^d_{23} & i s_1 \underline{h}^d_{12} \\
\star \ & \star & \underline{h}^d_{33}  & is_1 \underline{h}^d_{13}+ c_1 \underline{h}^d_{34}   \\
0 \ & \star & \star  & 0 \\
\end{array}\right)
\label{eq:Md_texture}
\end{equation}
where the $\star$ in the lower triangle is a short hand notation to indicate that the term is obtained from the corresponding element in the upper triangle by complex conjugation to make the matrix Hermitian. By comparing~(\ref{eq:Mu_texture}) and (\ref{eq:Md_texture}) with equations~(\ref{eq:Md_diag_basis})  and~(\ref{eq:Mu_Md_diag_basis})we can get the Yukawa's in the physical basis ($h^{u,d}_{ij}$) in terms of those in the symmetry basis ($\underline{h}^{u,d}_{ij}$). Now using~(\ref{eq:theta_F}) we find the contribution to $\bar{\theta}$ from radiative corrections of $M_u$ is:
\begin{eqnarray}
\bar{\theta}  \sim  {\frac{1}{16 \pi^2}}    \ Im \left(\frac{{h^{u}_{14}}{h^{d}_{43}}h^{d}_{31}}{h^u_{11}}\right) \ell \nonumber \\
\sim {\frac{1}{16 \pi^2}}  \left(\frac{{-s_1 \underline{h}^{u}_{11}}{\underline{h}^{d}_{34}}\underline{h}^{d}_{13}}{\underline{h}^u_{11}}\right) \ell
\label{eq:10^-11}
\end{eqnarray}
where we have dropped terms of order $s^2_1$ and taken $c_1 \sim 1$.  Since the light $3 \times 3$ up sector is diagonal in~(\ref{eq:Mu_texture}), we have $s_1 \underline{h}^d_{34} \sim V_{ub} \underline{h}^{d}_{33}$ and $\underline{h}^d_{13} / (s_1 \underline{h}^d_{34}) \sim cos \gamma$ where from the Particle Data Group's global CKM fits~\cite{0954-3899-37-7A-075021} we have $V_{ub} \sim 3.47 \times 10^{-3}$ and $\gamma \sim 68.8^o$ is approximately the CKM $CP$ phase. Substituting in~(\ref{eq:10^-11}) we get
\begin{equation}
\bar{\theta} \sim {\frac{1}{16 \pi^2}}  {\frac{m^2_b}{m^2_t}} V^2_{ub} cos\gamma \ell \sim   10^{-11} \ell
\end{equation}
where $\underline{h}^{d}_{33} = m_b/m_t \sim 1/40$ is the bottom quark Yukawa.

We similarly find for all choices of the placement of the  non-zero off diagonal elements that the strong $CP$ phase generated has the lower bound $\bar{\theta} \geq 10^{-11}$.  

While making the above statement we note that there is a special case with the choice  $\underline{h}^{u,d}_{i4} = 0$ in equation~(\ref{eq:massmatrixZ2}) for all $i= 1$ to $4$ as it can lead to a vanishing $\bar{\theta}$ at the one-loop level.  However  this texture does not generate the observed $CKM$ phase. With $M_i \neq 0$ in~(\ref{eq:massterms}) and~(\ref{eq:massmatrixZ2})  this can be easily seen by first rotating in the $1-4$ plane, then in the $2-4$ and  finally in the $3-4$ plane and going to the physical basis of~(\ref{eq:Md_diag_basis}) and~(\ref{eq:Mu_Md_diag_basis}).  Due to the hierarchy of the Yukawa couplings and Yukawas involving the first generation being the smallest, there is no significant mixing generated by the above rotations in the light $3 \times 3$ sector between Yukawa terms in the first row and those in the second and third rows and therefore the CKM phase generated is too small. This texture is thus not easily viable. 

An interesting thing about equation~(\ref{eq:10^-11}) is that the Yukawas involving the fourth heavy quarks such as $\underline{h}^d_{34}$ are multiplied by the angle $s_1$ so that the product leads to a Yukawa term of the light first generation. This means that $\underline{h}^d_{34}$ itself does not have to be small, it is the product $s_1 \underline{h}^d_{34}$ that is small. 

Radiative corrections to $M_d$ can also contribute but they turn out to be much smaller for  textures with maximal $CP$ such as those in~(\ref{eq:Mu_texture}) and~(\ref{eq:Md_texture}).  However they are relevant if the $CP$ phase is not maximal as we will see in the next section.

\subsection{Neutron EDM Bound for General CP Phase} 
\label{sec:nedmany}
If $CP$ is broken softly by dimension 3 quark mass terms (ie) by choosing  $M_i$ to be complex or spontaneously via a real singlet (both these cases are studied in reference~\cite{Kuchimanchi:2010xs}) then the $CP$ phase in the quark mass matrix in the symmetry basis can be non-maximal and arbitrary. We can then choose it to be close to the CKM phase $\gamma$.  This means that with just 3 non-zero off-diagonal Yukawa elements  in the symmetry basis (and $M_2=M_3=0$) we can obtain the needed CKM angles and the CKM $CP$ phase.  For such textures, since there is one less non-zero term than in~\ref{sec:nedmpi2},  we find that the $\bar{\theta}$ generated will be smaller by about two orders of magnitude and accordingly the lower bound is reduced.

To see this note that if the $CP$ phase $i$ in the first row of  equation~(\ref{eq:Md_texture}) is replaced by $e^{i\gamma'}$ with $\gamma'$ close to the CKM phase $\gamma$, then we can set $\underline{h}^d_{13}$ to zero and still obtain the needed CKM angles and $\gamma$.  However if $\underline{h}^d_{13} = 0$ then the contribution to $\bar{\theta}$ from equation(~\ref{eq:10^-11}) vanishes.  

On the other hand we find that the contribution to $\bar{\theta}$ from radiative corrections to $M_d$ remain non-vanishing.  Since the light $3 \times 3$ sector of $M_d$ is non-diagonal we need to use the basis independent  equation~\ref{eq:theta_F_inv} with $u \leftrightarrow d$ to estimate them.  We find  

\begin{eqnarray}
\bar{\theta}  \sim  {\frac{1}{16 \pi^2}}    \ Im \left(\frac{ h^{d}_{12}h^{d}_{23}h^{d}_{34}h^u_{14}h^u_{11}}{h_{d} h_{s}h_b}\right) \ell \nonumber \\
\sim {\frac{sin \gamma}{16 \pi^2}}  \left(\frac{ \underline{h}^{d}_{12}\underline{h}^{d}_{23}  \underline{h}^{d}_{34}s\underline{h}^{u^2}_{11}}{h_{d} h_{s}h_b}\right) \ell
\label{eq:10^-13}
\end{eqnarray}
where the product of the Yukawas in the denominator corresponds to the determinant of the light $3 \times 3$ down-sector Yukawa matrix, that is $h_{d, s, b} = m_{d, s, b} / m_t$.  Substituting $\underline{h}^d_{12} = V_{us} \underline{h}^d_{22}, \ \underline{h}^d_{23} = V_{cb} \underline{h}^{d}_{33}$ and as before  $s \underline{h}^d_{34} = V_{us}\underline{h}^d_{33}$ we get
\begin{equation}
\bar{\theta} \sim {\frac{1}{16 \pi^2}} \left(\frac{m^2_u m_b}{m_d m^2_t}\right)V_{us}V_{cb}V_{ub} sin \gamma \ell \sim 10^{-13.5} \ell
\label{eq:13.5}
\end{equation}

There is one more potential source of radiative corrections at the one-loop level that is relevant.  In order to calculate lower bounds we have assumed textures with zeros in mass matrices  which are not necessarily supported by any symmetries and therefore small non-zero values will be radiatively generated.  To avoid fine-tuning, the natural tree-level values for these elements must be greater than the one-loop corrections.

Since the bi-doublets will acquire VEVs diag $\{\kappa, \kappa'\}$ if a matrix element $\underline{h}^d_{ij}$ is non-zero then the corresponding element $\underline{h}^u_{ij}$ must be at least of the order $\underline{h}^d_{ij}\kappa'/\kappa$ (with $\kappa' << \kappa$)  since both the up and down sector can receive contribution from a term such as $\underline{\tilde{h}} \bar{Q}_{iL}\tilde{\phi} Q_{jR}$.  The same is also true if the up sector has a non-zero element the corresponding down element will also get a small VEV by the same argument.

\begin{figure}[ht]  
\includegraphics[width=45mm]{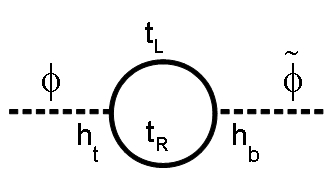}
\caption{Radiative CP invariant contribution to $\kappa'$}
\label{fig:kappa_prime}
\end{figure}

Even if $\kappa'$ is set to zero it will acquire a one loop correction order $\kappa /(16 \pi^2) (m_b/m_t) \sim \kappa/6000$ due to the diagram in figure~\ref{fig:kappa_prime} where in the figure $h_t= \underline{h}_{33}, \ h_b = \underline{\tilde{h}}_{33}$ are the top and bottom quark Yukawas.
 Thus instead of zeros we should have a value for every element in the up (down) Yukwas that is at least $1/6000^{th}$ of the corresponding element in the down (up) Yukawas. Since this is very small it doesn't increase the lower bounds we have evaluated in~(\ref{eq:13.5}).

However where this is relevant is for textures that may have an additional zero in the diagonal element such as:
\begin{equation}
H_u =
\left(\begin{array}{cccc}      
0 & 0  & c_1 e^{i\gamma'}\underline{h}^u_{13}   & 0  \\       
0  & \underline{h}^u_{22}  & \underline{h}^u_{23} & 0 \\
\star &   \star  & \underline{h}^u_{33}  & -s_1 e^{-i\gamma'}\underline{h}^u_{13}\\
0 & 0 & \star  & 0 \\
\end{array}\right)
\label{eq:Mu_RRR}
\end{equation}
\newline
\noindent
\begin{equation}
H_d = 
\left(\begin{array}{cccc}      
0 \ & s_1 \underline{h}^d_{24}   & c_1\epsilon  & 0 \\    
\star \ & \underline{h}^d_{22}  & 0&  c_1 \underline{h}^d_{24} \\
\star \ & 0 & \underline{h}^d_{33}  & -s_1 \epsilon ^\star   \\
0 \ & \star & \star  & 0 \\
\end{array}\right)
\label{eq:Md_RRR}
\end{equation}
The light $3 \times 3$ sub-matrices of the above have a predictive texture~\cite{Ramond:1993kv} and for $\epsilon=0$ it is known in the literature as
the texture corresponding to solution 5 of Ramond, Roberts and Ross (RRR). This texture gives the approximate predictions $V_{us} = \sqrt{m_d/m_s}$ and $V_{ub} = \sqrt{m_u/m_t}$.  If we now evaluate $\bar{\theta}$ using the above matrices we find that the contribution from radiative corrections to $M_u$ as well as $M_d$ vanishes if $\epsilon = 0$. Thus it maybe possible that the lower bound may be weaker than~(\ref{eq:13.5}) for the case where CP violation is non-maximal, but we eliminate this possibility below.

As we have argued $\epsilon$ cannot be taken to be zero but must at least be $\epsilon \sim e^{i\gamma'}\underline{h}^u_{13} /6000$ so as to prevent fine-tuning.  Evaluating the contribution to $\bar{\theta}$ from radiative corrections to $M_d$ by using equation~(\ref{eq:theta_F_inv}) with $u \leftrightarrow d$ we get for the above $\epsilon$
\begin{eqnarray}
\bar{\theta} \sim & & \frac{\displaystyle \sin(2\gamma')}{\displaystyle 16 \pi^2} \frac{\displaystyle s^2_1 \underline{h}^{d^2}_{24} \underline{h}^{u^2}_{13} \underline{h}^u_{33}}{\displaystyle 6000 \ h_d h_s h_b} \ell \nonumber \\ \sim & &\frac{\displaystyle \sin(2\gamma')}{\displaystyle 96000 \pi^2} (\displaystyle V^2_{us}V^2_{ub}) {\frac{\displaystyle m_s m_t}{\displaystyle m_d m_b}} \ell \nonumber \\ \sim & &3\times 10^{-10} \ell
\end{eqnarray}
where we used $(s_1 \underline{h}_{24}) / h_s = V_{us}, \ \underline{h}^u_{33} \sim h_t, \ \underline{h}^u_{13}/\underline{h}^u_{33} \sim V_{ub}$ and $h_{d, s,b,t} = m_{d, s, b,t} / m_t$ with the central values $ V_{us} = 0.2253, \ V_{ub} = 0.00347, \ m_d = 5.1 MeV, \ m_s = 100 MeV, \ m_b = 4.19 GeV$ and $m_t = 173 GeV$.

Note that the above quantity came out higher than the previous cases because off-diagonal Yukawas in the up quark sector that give rise to any CKM mixing angles are $10$ to $40$ times higher than if the CKM angles resulted from a down quark Yukawa.  And therefore $\bar{\theta}$ is comparatively higher if up sector Yukawas are involved in generation of CKM mixing angles.
        
Thus we find that the lowest value we can get for the case of general (non-maximal) $CP$ violation is given by~(\ref{eq:13.5}) and this serves as the lower bound.  

We have thus shown that for $M_{H^+_2} > M$ the lower bound for $\bar{\theta}$ is $10^{-13}$ for non-maximal $CP$ violation and $10^{-11}$ for maximal CP violation.  These bounds correspond to $d_n \geq 10^{-29} ecm$ and $\geq 10^{-27} ecm$ respectively.  For $M_{H^+_2} < M$ there is a further suppression by factor  $(M_{H^+_2}/M)^2$
 
We now see to what extent we can predict  the CKM phase that can arise from the maximal phase.

\section{CKM phase prediction from texture and flavour}
\label{sec:CKM}

Since a maximal $CP$ phase is generated in the symmetry basis it is tempting to see if  textures and flavour symmetries can help predict the CKM phase that is obtained from the maximal phase.   
\subsection{Using Texture}
\label{sec:texture}
It is well-known that setting some of the diagonal elements of Yukawa matrix to zero will lead to a predictive mass-mixing angle relation, as the mass would then have to come from an off-diagonal element.   We can use a similar trick to predict the CKM phase from texture alone.

As an example we take  $\underline{h}^{u,d}_{11} =0$  and take $M_2, M_3 = 0$ and take the four   non-zero  off diagonal Yukawas to be $\underline{h}^{d}_{12}, \underline{h}^{d}_{23}, \underline{h}^{d}_{34}$ and $\underline{h}^u_{24}$.  We follow the same procedure as in  section~\ref{sec:nedmpi2}.

Diagonalizing in the 1-4 plane so that the heavy quarks  decouple (with $s = M_1 / M$ as before), we get the $4\times 4$ Yukawa matrices in the physical basis to be:

\begin{equation}
H_u =
\left(\begin{array}{cccc}      
0 & s \underline{h}^u_{24}  & 0   & 0  \\       
\star  & \underline{h}^u_{22}  & 0 &c \underline{h}^u_{24} \\
0 & 0   & \underline{h}^u_{33}  & 0 \\
0 & * & 0  & 0  \\
\end{array}\right)
\label{eq:Mu_texture_CP}
\end{equation}
\newline
\noindent
\begin{equation}
H_d = 
\left(\begin{array}{cccc}      
0 \ & i c \underline{h}^d_{12}   & s \underline{h}^d_{34}  & 0 \\    
\star \ & \underline{h}^d_{22}  & \underline{h}^d_{23} & i s \underline{h}^d_{12} \\
\star \ & \star & \underline{h}^d_{33}  &  c \underline{h}^d_{34}   \\
0 \ & \star & \star  & 0 \\
\end{array}\right)
\label{eq:Md_texture_CP}
\end{equation}
We can see from the top $2 \times 2$ sub-matrices in the above two matrices that we get the predictive relations $m_d  \sim V^2_{us} m_s$ (from~(\ref{eq:Md_texture_CP}) with $\underline{h}^d_{22} v = m_s$ and $V_{us} = c\underline{h}^d_{12} /  \underline{h}^d_{22}$) , and $cos \gamma \sim (\sqrt{m_u/m_c}) / V_{us}$ (from~(\ref{eq:Mu_texture_CP}) with $\underline{h}^u_{22} v = m_c$ and $m_u = s^2 \underline{h}^{u^2}_{24} / \underline{h}^u_{22}$ and from ~(\ref{eq:Md_texture_CP})) . Thus we get $\gamma \sim 79^o$ where we have used the central values $m_u = 2.5 MeV, \ m_c = 1290 MeV, \ V_{us} = 0.225$.  

If we diagonalize the full $3 \times 3$ light quark sub-matrix and do the calculation more accurately  $\gamma$ can be a bit lowered to $76^o$.  Comparing this with the experimental value of $\gamma = 68.8^o \pm 3^o$ we see that while the agreement is not too great, the texture does come pretty close and gives an idea of how the CKM matrix may get generated from the maximal phase.  Of course once we turn on the terms we set to zero we can easily get corrections to the texture prediction and an agreement with experiments.

We  evaluate the $\bar{\theta}$ for this texture and find it to be 
\begin{equation}
\bar{\theta} = \frac{ \sin \gamma}{16\pi^2} \frac{s h^d_{12} h^u_{24} m_s}{m_c} \ell \sim 10^{-11} \ell
\end{equation}

We will now see how close we can get to predicting the CKM phase using flavour symmetries instead of just texture. In the next sub-section we motivate a $Z_4$ symmetry and evaluate how the maximal CP phase and $Z_4$ symmetry can help understand how the CKM phase may be generated.  
   
\subsection{CKM Phase Prediction Using $Z_4$ Flavour}
\label{sec:Z4}
If we take seriously the observed fact that the down quark mass $m_d$ is related to the Cabibo angle via $m_d = V_{us}^2 m_s$, the two conditions that are needed to obtain  this relation are:

\begin{itemize}
	\item Quark Mass matrix is Hermitian.
\item The down quark mass matrix has a near zero in its first diagonal entry.
\end{itemize}

Our strong $CP$ solution  ensures Hermiticity of the quark mass matrices  and we impose an additional $Z_4$ symmetry to ensure the second condition. 

Once $Z_4$ is imposed $\phi_s$ and $\tilde{\phi_s}$ will transform differently under it and therefore both of them will not couple to the same quark pairs. This means that ignoring off-diagonal elements, the quark mass ratios will be the same for all generations since every generation would get its mass from  $\phi_s$ and not from $\tilde{\phi}_s$. Therefore if we fix $\kappa'_s / \kappa_s = m_b / m_t \sim 1/40 $ then since $m_c /40 $ and $m_u/40$ are much smaller than $m_s$ and $m_d$, they would have to get their masses from off-diagonal terms. These constraints can lead to the predictions of CKM $CP$ phase,  and as expected the down quark mass.   

Under $Z_4$ we have 
\begin{eqnarray}
& Q_{1L,4L}, Q'_{R} \rightarrow i Q_{1L,4L}, iQ'_{R}; \ \ Q_{2L,3L} \rightarrow - i Q_{2L,3L}; \nonumber \\& Q_{2R,3R} \rightarrow - Q_{2R,3R}; \ \ \phi_s, \phi_a \rightarrow i\phi_s, i\phi_a
\end{eqnarray}
while the rest of the quark and Higgs fields are invariant. The leptons can also transform non-trivially under $Z_4$ but we do not write these out explicitly.

The most general Yukawa potential invariant under $P \times CP \times Z_2 \times Z_4$ is:
%\begin{widetext}
%\begin{equation}
\begin{eqnarray}
\sum_{i= 1~to~4}& \underline{h}_{ii} \bar{Q}_{iL} \phi_s {Q}_{iR} + \underline{h}'\bar{Q}'_L \phi^{\dagger}_s Q'_R + \nonumber \\ \sum_{j=2~to~3}  & [i\underline{\tilde{h}}_{1j} (\bar{Q}_{1L} \tilde{\phi}_a {Q}_{jR}   -  \bar{Q}_{jL} \tilde{\phi}_a {Q}_{1R}) \nonumber \\ & +  \underline{\tilde{h}}_{j4}  (\bar{Q}_{jL} \tilde{\phi}_s {Q}_{R4} +  \bar{Q}_{L4} \tilde{\phi}_s {Q}_{jR})]  \nonumber \\ + i \underline{h}_{14} & (\bar{Q}_{1L} {\phi}_a {Q}_{R4} - \bar{Q}_{L4} {\phi}_a {Q}_{1R})   + h.c. 
\label{eq:Z4Yukawa}
\end{eqnarray}
%\end{equation}
%\end{widetext} 
where without loss of generality we have chosen a basis such that $\underline{\tilde{h}}_{23} = 0$.

Once the diagonal elements of $\phi_{a,s}$ pick up real VEVs diag$\{\kappa_{a,s}, \kappa'_{a,s}\}$ the mass matrix $M_d$ is given by
\begin{equation}
M_d =
\left(\begin{array}{ccccl}      
\underline{h}_{11} \kappa'_s  & i\underline{\tilde{h}}_{12} \kappa_a & i\underline{\tilde{h}}_{13} \kappa_a & i \underline{h}_{14} \kappa'_a &  M_1 \\       
-i\underline{\tilde{h}}_{12} \kappa_a  & \underline{h}_{22} \kappa'_s  & 0 &  \underline{\tilde{h}}_{24} \kappa_s&  M_2 \\
-i\underline{\tilde{h}}_{13} \kappa_a  & 0  & \underline{h}_{33} \kappa'_s &  \underline{\tilde{h}}_{34} \kappa_s & M_{3} \\
-i \underline{h}_{14} \kappa'_a&  \underline{\tilde{h}}_{24} \kappa_s  &  \underline{\tilde{h}}_{34} \kappa_s & \underline{h}_{44} \kappa'_s& M_{4} \\
M_1 & M_2 & M_{3} & M_{4} &  \underline{h}' \kappa'_s 
\end{array}\right)
\label{eq:massmatrixZ4_Md}
\end{equation}

$M_u$ is given by a similar matrix with $\kappa_{a,s} \leftrightarrow \kappa'_{a,s}$.  Note that since we set $\kappa'_s / \kappa_s = m_b/m_t$, $M_u$ is almost diagonal with negligible off-diagonal elements while the first two diagonal elements of $M_d$ are small and need contribution from off-diagonal terms to provide the right masses for down and strange quarks.      

When we make an orthogonal transformation in the $2-4$ plane by an angle $s_2 \sim M_2 / \sqrt{M_2^2 + M_4^2}$ to go to the physical basis,  the strange quark mass picks up an additional contribution so that it becomes 
\begin{eqnarray}
m^o_s \sim &\underline{h}_{22} \kappa'_s + 2 s_2 \underline{\tilde{h}}_{24} \kappa_s \nonumber \\ \sim &\frac{\displaystyle m_c m_b}{\displaystyle m_t} + 2 s_2 \underline{\tilde{h}}_{24} \kappa_s \nonumber \\ \sim & \pm \left|\frac{\displaystyle m_c m_b}{\displaystyle m_t} \right| + 2 s_2 \underline{\tilde{h}}_{24} \kappa_s
\label{eq:m_s}
\end{eqnarray}
where in the last line we have a $\pm$ since $m_c = \underline{h}_{22} \kappa$  and can be positive or negative depending on the sign of $\underline{h}_{22}$.

To keep matter simple we set $M_3 =0, \ \underline{h}_{14} =0$ and we rotate along the $1-4$ plane by an angle $s_1 = M_1/\sqrt{{M_1^2 + M_2^2 + M_4^2}}$.  This takes us to the physical basis and the light $3 \times 3$ sector of $M_d$ becomes:
\begin{equation}
M_{d3\times 3} =
\left(\begin{array}{ccc}      
0 & i\underline{\tilde{h}}_{12} \kappa_a + s_1 \underline{\tilde{h}}_{24} \kappa_s & i\underline{\tilde{h}}_{13} \kappa_a + s_1 \underline{\tilde{h}}_{34} \kappa_s \nonumber \\       
\star  & m^o_s  & s_2 \underline{\tilde{h}}_{34} \kappa_s \nonumber \\
\star  & \star  & m_b 
\end{array}\right)
\label{eq:massmatrixZ4_Md33}
\end{equation}
We will first do an approximate calculation. If $\underline{h}_{13} << s_1\underline{h}_{34}$ then from~(\ref{eq:massmatrixZ4_Md33}) we get
\begin{equation}
s_1 / s_2 \sim \left|V_{ub} / V_{cb}\right| 
\label{eq:s1s2}
\end{equation}
and using equation~(\ref{eq:m_s}) with~(\ref{eq:s1s2}) we get
\begin{equation} 
s_1 \underline{\tilde{h}}_{24} \kappa_s \sim \left|\frac{V_{ub}}{\displaystyle 2V_{cb}}\right| \left(m^o_s \mp \left|\frac{m_c m_b}{\displaystyle m_t} \right|\right)
\end{equation}
Substituting the above in~(\ref{eq:massmatrixZ4_Md33}) we get for the phase in the down quark mass matrix
\begin{eqnarray}
\gamma^o \sim & \cos^{-1}&\frac{\displaystyle s_1 \underline{\tilde{h}}_{24} \kappa_s}{\displaystyle V_{us}m^o_s} \nonumber \\\nonumber \\ \sim & \cos^{-1} & \left[\left|\frac{\displaystyle V_{ub}}{\displaystyle 2V_{cb}V_{us}}\right| \left(1 \mp \left|\frac{\displaystyle m_c m_b}{\displaystyle m^o_s m_t} \right|\right)\right] \sim 75.7^o
\end{eqnarray}
where we have used from the particle data group's global fits to CKM matrix~\cite{0954-3899-37-7A-075021} $V_{us} = 0.2253, V_{cb} = 0.0410, V_{ub} = 0.00347$ and $m^0_s \sim m_s = 100 MeV, m_c = 1.290 GeV, m_b = 4.19 GeV$ and $m_t = 173 GeV$. We have used the $+$ from $\mp$ to get the lower value of $\gamma^o$.

The phase $\gamma^o$  in the mass matrix $M_d$ (in basis with $M_u$ diagonal) is related to the CKM phase $\gamma$ roughly by the relation $\gamma = \gamma^o - 3.4^o$ and so we find that due to $Z_4$ symmetry the maximal phase $\pi / 2$ results in a CKM phase 
\begin{equation}
\gamma = 72^o \pm \Delta \left(\gamma\right)_{\underline{\tilde{h}}_{13}}
\label{eq:gamma}
\end{equation}
where 
\begin{equation}
\Delta \left(\gamma\right)_{\underline{\tilde{h}}_{13}} = \sin^{-1} \frac {\underline{\tilde{h}}_{13}\kappa_a}{m_b V_{ub}}
\label{eq:deltagamma}
\end{equation}
is the contribution due to a small non-zero  $\underline{\tilde{h}}_{13}$.  

There is a subtle point  in the above equations.  In principle $\underline{\tilde{h}}_{13}$ can also contribute to change the first term in~(\ref{eq:gamma}) from $72^o$ to a different number.  This is because the exact value depends on equation~(\ref{eq:s1s2}) which will get modified once $\underline{\tilde{h}}_{13}$ is non-zero to
\begin{equation}
\frac{\sqrt{(\underline{\tilde{h}}_{13}\kappa_a)^2 + (s_1 \underline{\tilde{h}}_{34} \kappa_s)^2}}{s_2 \underline{\tilde{h}}_{34}\kappa_s} \sim \left| \frac{V_{ub}}{V_{cb}}\right| 
\end{equation}
Expanding for small $\underline{\tilde{h}}_{13}$ we get
\begin{equation}
s_1 / s_2 + O(\underline{\tilde{h}}^2_{13}) = V_{ub} / V_{cb}
\end{equation}
and thus there is no $O(\underline{\tilde{h}}_{13})$ correction to the first term of equation~(\ref{eq:gamma}) which remains $72^o$.  Equation~(\ref{eq:gamma}) with~(\ref{eq:deltagamma}) is thus correct to $O(\underline{\tilde{h}}^2_{13})$.

The other Yukawa we set to zero $\underline{h}_{14}$ can be treated analogously and like $\underline{\tilde{h}}_{13}$ it does not change the $72^o$ but just makes an additive contribution to the lowest order when its turned on. 

Thus we see that the maximal phase generated in the quark mass matrix is actually  changed to $72^o$ on diagonalization to the physical mass basis due to known terms when $Z_4$ is imposed, while there can be a small correction from this value by unknown Yukawa terms that can be set to zero without affecting anything else.  The value $72^o$ is obtained by inputting central values from experiments for quark masses and CKM parameters (other than the CKM CP phase) and therefore it can have an experimental error of about $3^o$.  This is thus consistent with the current experimentally determined value $\gamma = 68.8^o \pm 3^o$ even if Yukawas like $\underline{\tilde{h}}_{13}$ were too small to contribute any amount.

There are a couple of points we need to make before we proceed.  Equations like ~(\ref{eq:s1s2}) that relate CKM parameters to the symmetry basis Yukawas have been derived  approximately so they can be presented simply but we can do a more exact derivation by using CKM matrix to $O(\lambda^4)$ in Wolfenstein parametrization.   The resulting equation~(\ref{eq:gamma}) remains the same to within a degree when we do this.  We also had set $M_3 \sim 0$ and when it is turned on it will also make a small contribution to equation~(\ref{eq:gamma}) and so must be kept small. However including $M_3$ can also significantly contribute to $\bar{\theta}$ and $M_3$ can be bounded  based on that.  

We now evaluate  $\bar{\theta}$ for this case using~(\ref{eq:theta_F}).  We find
the dominant contribution is due to radiative corrections to $M_u$ and is
\begin{eqnarray}
\bar{\theta} \sim & \left( \frac{\displaystyle 1}{\displaystyle 16\pi^2}\right)\left( \frac{\displaystyle \kappa_a}{\displaystyle \kappa_s}\right) s_1 \underline{\tilde{h}}_{12} \underline{\tilde{h}}_{24}  \ell \nonumber \\ \sim & \left( \frac{\displaystyle \cos \gamma}{\displaystyle 16\pi^2} \right) \frac{\displaystyle V_{us}^2 m^2_s}{\displaystyle m^2_t} \ell \nonumber \\ \sim &  3.5 \times 10^{-11} \ell
\end{eqnarray}

We have used $Z_4$ symmetry as an example to show how the CKM phase and $d_n$ may be calculated based on  flavour symmetries.  Other flavour symmetries can be similarly explored.  Especially interesting would be those that are also motivated from the lepton sector considering the neutrino mixing.  

\section{Few Comments}
\label{sec:comments}
Before we conclude we make some comments:

\begin{enumerate}
\item In this work we have shown how P and CP symmetry breaking can lead to predictions for the neutron's electric dipole moment. For a large region of 
parameter space of the model these predictions turn out to be in the region that will be experimentally probed in the next few years. These experiments~\cite{Burghoff:2011xk,*Baker:2010,*Beck:2011gw,*Martin:2011,*Serebrov:2009zz} are often considered to be complimentary to the LHC and are currently facing various hurdles~\cite{2012Nature}. However since the neutron EDM generated due to $P$ and $CP$ breaking does not get diminished even if there is no new physics up until the see-saw, GUT or Planck scales these experiments can in fact make a discovery due to physics  at scales that are out of reach for LHC or future colliders. The last remaining parameter of the standard model $\bar{\theta}$ that gives rise to  $d_n$, as well as  neutrino masses and mixing,  could both be a consequence of very high energy physics emerging from the left-right symmetric strong CP solving model. 

	\item Once CP is imposed, it will also apply on the lepton sector. If there are no vectorlike leptons (analogous to the vectorlike quarks) then $CP$ phase is not generated at the tree level in the lepton sector, and there will be no $CP$ phase in the neutrino mixing parameters.  If there are vectorlike leptons then the same mechanism of $CP$ generation can result in a maximal phase in the symmetry basis of the lepton sector (if $CP$ is broken spontaneously by a complex singlet or softly by dimension 2 terms) or a general CP phase (if it is broken by real singlet or softly by dimension 3 fermion mass terms).  If there is Grand Unification, then we would expect vectorlike leptons to also be present.
	
\item Our model is generalizable to the supersymmetric left-right model (SUSYLR).  In fact if we want to break  $CP$ spontaneously in SUSYLR so as to reduce the number of CP phases such as from the gaugino sector~\cite{Kuchimanchi:1995rp,*PhysRevLett.76.3490,*Mohapatra:1997su,*Pospelov:1996be} that could generate too large a $d_n$, then we can introduce the vector-like quarks as we have done with similar consequences. 

\item We have used a complete vectorlike quark family with $SU(2)_L$ doublet and parity related iso-singlet ($SU(2)_R$ doublet) quarks. In fact if only the Higgs triplets $\Delta_{L,R}$ that are the natural candidates for providing large Majorana masses to the right-handed neutrinos are used to break $SU(2)_R$, then this is the only choice of vectorlike quarks that works for solving the strong CP problem, since only they can have Yukawa couplings with bi-doublets and mix with the lighter quarks. So these quarks are uniquely picked in some sense.  
 
\item  The vectorlike quark masses are protected from quadratic divergences from the heavy sector since in the model without scalar singlets they only couple to the standard model Higgs VEVs (through Yukawa couplings with the bi-doublets) and through dimension 3 fermion mass terms. Thus these quarks can be naturally light including being at scales accessible by LHC and future colliders as pointed in our previous work~\cite{Kuchimanchi:2010xs}.  Of course the dimension 3 mass terms can also be very heavy like being at the GUT scale.   Their mass is an independent scale and can be light or heavy. 
\end{enumerate}
 
\section{Concluding Remarks}
\label{sec:conclusions}
1950's and 60's presented the surprising discoveries that nature through weak interactions violates parity ($P$) and matter-antimatter ($CP$) symmetries. However while doing so nature has left us a puzzling situation of an extremely small CP violation by the strong interactions (as yet undetected) that seems to imply that there is an unknown hidden symmetry, along with a large CKM phase that seems to be put in by hand  in the form of hard CP violation in the Yukawa and weak sector. If the unknown symmetries are in fact $P$ and $CP$ themselves which are spontaneously or softly broken and therefore hidden from our view, then we have shown in this work that the neutron must know about it  since its electric dipole moment gets generated when these are broken. 

Finding a neutron EDM consistent with the predictions in this work can provide evidence for  $P$ and $CP$ symmetric laws of nature that would have been visible at high energies  such as those at the time of the origin of the universe..

Moreover we find that the CP phases of Higgs VEVs that are generated spontaneously  get determined by the $P$ transformation properties since the strong CP solving vacuum is such that VEVs that violate $CP$ must conserve $P$. This implies that for Higgses that couple to quarks and leptons the $CP$ phases of the VEVs are maximal.  

In addition if $CP$ violation happens either softly by dimension 2 terms or  spontaneously with a complex singlet we find that the $CP$ phase generated in the quark (in general fermion) mass matrix is also maximal in the symmetry basis, thereby providing an understanding of the largeness of the CKM phase. The same symmetries such as $Z_2$ that are needed to ensure that there is no tree-level strong CP phase determine the maximality of the weak CP phases in the quark (fermion) mass matrix.  While if $CP$ is broken softly by dimension 3 fermion mass terms or spontaneously by a real singlet then  the tree-level strong CP phase is automatically not present and no further symmetries like $Z_2$ are needed and as a result the CP violation in the quark (fermion) mass matrix need not also be maximal. In both cases vectorlike quarks (leptons) are needed to ensure the $CP$ violation in the quark (lepton) matrices cannot trivially be rotated away.

The strong $CP$ phase is radiatively generated at the one-loop level due to the $SU(2)_L$ breaking charged would be goldstone bosons and can be estimated from their Yukawa couplings with the quarks that include a full vectorlike quark family. $P$  ensures that there is a mode in the Higgses of the right handed sector that cancels the divergences from the goldstone mode and a finite amount of $\bar{\theta}$ is generated that depends logarithmically on the mass ratio of the heavy higgs $M_{H^+_2}$ and the heavy quark mass $M$ for the case where $M_{H^+_2} < M$.    

The results for the $\bar{\theta}$ and $d_n$ generated are:

\begin{itemize}
	\item \bfseries For case of maximal CP violation: \normalfont \newline 
	\begin{equation}
	\bar{\theta} \geq 10^{-11} \times
   \left\{
  \begin{array}{l l}
    \ln(M_{H^+_2}/M) \geq 1 & \quad for \ M < M_{H^+_2} \nonumber\\
   (M_{H^+_2}/M)^2  & \quad for \ M > M_{H^+_2} \\
  \end{array} \right.
  \label{eq:conclusion}
\end{equation}
\item \bfseries For case of general CP violation: \normalfont \newline 

\begin{equation}
	\bar{\theta} \geq 10^{-13} \times
   \left\{
  \begin{array}{l l}
    \ln(M_{H^+_2}/M) \geq 1 & \quad for \ M < M_{H^+_2} \nonumber\\
   (M_{H^+_2}/M)^2  & \quad for \ M > M_{H^+_2} \\
  \end{array} \right.
  \label{eq:conclusion2}
\end{equation}
\end{itemize}
and $d_n$ can be obtained from the relation $d_n \sim 2 \times 10^{-16} \bar{\theta} \ ecm$~\cite{Pospelov:2005pr}.

If we are fortunate we will be able to observe the neutron EDM in the ongoing searches~\cite{Burghoff:2011xk,*Baker:2010,*Beck:2011gw,*Martin:2011,*Serebrov:2009zz} that  plan to probe $d_n \sim 10^{-26}$ to $10^{-27} ecm$ in the next few years and up to $10^{-29} ecm$ in the coming two decades.   A lot depends on whether $M < M_{H^+_2}$ or if its larger, then by how much? We know that the heavy doublet Higgs mass $M_{H^+_2}$ must be around the right-handed symmetry (parity) breaking scale. Based on the smallness of the neutrino mass this would most naturally be at or above the see-saw scale of $10^{14} GeV$ if we assume that the neutrino generations have similar Yukawa coupling with the standard model Higgs like their counterpart up sector quark generations.  We also know that $M$ being a fermion mass term is chirally protected and can be naturally small. Therefore there are reasonably good chances to observe the electric dipole moment of the neutron in the planned experiments that could provide evidence for the hidden $P$ and $CP$ symmetries in nature.

Equations~(\ref{eq:conclusion}) and~(\ref{eq:conclusion2}) can also be used in the reverse so as to provide bounds for the mass ratio of the heavy higgs and heavy quarks based on non-observation of $d_n$.  For example the experimental bound $\bar{\theta} < 10^{-10}$ implies from~(\ref{eq:conclusion}) for the case of maximal $CP$ violation that the vectorlike quarks mass must be $M > e^{-10} M_{H^+_2} \sim M_{H^+_2}/20000$ so that the logarithm does not contribute more than a factor of $10$. This together with $M_{H^+_2} \sim 10^{14} GeV$ would imply that such vectorlike quarks would be out of reach of the LHC if CP violation is maximal. However some recent work shows that the current experimental bound on $\bar{\theta}$ could be a factor of 5 weaker~\cite{Narison:2008jp} based on the methods used to obtain the relationship between $d_n$ and $\bar{\theta}$. If this is the case there is still a chance for the vectorlike quarks with maximal CP violation to be at the mass scales being probed by the LHC. Of course for general, non-maximal $CP$ violation there is as yet no relevant constraint on the scale of the vectorlike quarks masses from neutron EDM experiments and they could be within the LHC reach or at very high scales. 
\begin{widetext}
\appendix
\section{Scalar potential With $P \times CP \times Z_2$ \\ and 2 Bi-doublets}
\label{sec:appendix}

We write down the Higgs potential in terms of the $P \times CP \times Z_2$ invariant part $V_{Inv}$, and soft $CP$ and $Z_2$ breaking part $V_{soft}$ that conserves $P$.
We note from equations~(\ref{eq:Vhiggs}) and~(\ref{eq:Vinva}) that
$$V_{Higgs} = V_{Inv.} + V_{soft}$$ 

with $$V_{Inv.} = V_{tri}(\Delta_L, \Delta_R) + V_{bi}(\phi_s, \phi_a)+ \sum_{z=a, s}{V(\phi_z, \Delta_L, \Delta_R)}$$
where $V_{tri}(\Delta_L, \Delta_R) + V(\phi_z, \Delta_L, \Delta_R)$ is the most general potential with 1 bi-doublet $\phi_z$ as given in reference~\cite{PhysRevD.44.837}.
%\begin{widetext}
\begin{eqnarray}
V_{tri}(\Delta_L, \Delta_R) = &  \{ -\mu_3^2 Tr(\Delta_R \Delta_R^\dagger) +\rho_1 Tr [(\Delta_R \Delta_R^\dagger)]^2 + \rho_2 Tr(\Delta_R \Delta_R)Tr(\Delta_R^\dagger \Delta_R^\dagger) +  \rho_4 Tr (\Delta_L \Delta_L) Tr(\Delta_R^\dagger \Delta_R^\dagger) + R \leftrightarrow L\} \nonumber \\
 \ & + \rho_3 Tr (\Delta_L \Delta_L^\dagger) Tr (\Delta_R \Delta_R^\dagger)
 \label{eq:A1} 
\end{eqnarray}
\begin{eqnarray}
V(\phi_z, \Delta_L, \Delta_R) = & -\mu_{1z}^2 Tr (\phi_z^\dagger \phi_z) -\mu_{2z}^2 [Tr(\tilde{\phi}_z^\dagger {\phi}_z) + h.c.] + \lambda_{1z} [Tr (\phi_z^\dagger \phi_z)]^2 + \lambda_{3z} Tr(\tilde{\phi_z}\phi_z^\dagger) Tr(\tilde{\phi_z}^\dagger\phi_z) +\{\lambda_{2z} [Tr(\tilde{\phi_z}\phi_z^\dagger)]^2 \nonumber \\ \ & + \lambda_{4z} Tr(\phi_z \phi_z^\dagger) Tr (\tilde{\phi}_z\phi_z^\dagger) + \beta_{1z}Tr(\phi_z \Delta_R \phi_z^\dagger \Delta_L^\dagger) + \beta_{2z}Tr(\tilde{\phi}_z \Delta_R \phi_z^\dagger \Delta_L^\dagger) + \beta_{3z}Tr(\phi_z \Delta_R \tilde{\phi}_z^\dagger \Delta_L^\dagger) + h.c.\} \nonumber \\ \ & +\{\alpha_{1z}Tr(\phi_z \phi_z^\dagger) Tr (\Delta_R \Delta_R^\dagger) + R\rightarrow L\} + \{\alpha_{2z}Tr(\phi_z \tilde{\phi}_z^\dagger) Tr (\Delta_R \Delta_R^\dagger) + R\rightarrow L + h.c.\} \nonumber \\ \ &+ \alpha_{3z} [Tr(\phi_z^\dagger \phi_z \Delta_R \Delta_R^\dagger) + Tr(\phi_z \phi_z^\dagger \Delta_L \Delta_L^\dagger)]
\label{eq:beta}
\end{eqnarray} 
\begin{eqnarray}
V_{bi}(\phi_a, \phi_s) = \lambda_{1as} Tr(\phi_a^\dagger \phi_a) Tr (\phi_s^\dagger \phi_s)  + \lambda_{2as} Tr(\phi_a \phi_a^\dagger \phi_s \phi_s^\dagger) + \lambda_{3as} Tr(\phi_s\tilde{\phi}_a^\dagger) Tr(\phi_s \tilde{\phi}_a^\dagger) +... 
\label{eq:A3}
\end{eqnarray}
\begin{eqnarray}
V_{soft} = -\mu_{1as}^2 Tr (\phi_a^\dagger \phi_s) - \mu_{2as}^2 Tr(\tilde{\phi}_a^\dagger {\phi}_s) - \mu_{2sa}^2 Tr(\tilde{\phi}_s^\dagger {\phi}_a)+ h.c.
\label{eq:A4}
\end{eqnarray}
 \end{widetext}
In the above we have used curly brackets so that terms that can be obtained by replacing $R$ by $L$ or those that can be obtained by Hermitian conjugation need not be independently written.  Within each curly bracket the operations such as $R \rightarrow L$ or $h.c.$ if mentioned would apply to every term and they generate the remaining terms.
  
 Note also that all the parameters in the above potential are real.  $\alpha_{2a}, \alpha_{2s}$ are real due to CP, and the parameters in $V_{soft}$ are real due to $P$.  The remaining parameters are all real due to $P$ and they are also real due to $CP$. Either of these symmetries can be used to make the remaining parameters real.  $Z_2$ ensures there are no other terms whose parameters  could be non-real.  Together these symmetries ensure that all parameters of the Higgs potential are real.
 
 $V_{bi}$ is made of several terms that necessarily involve both $\phi_a$ and $\phi_s$.  $Z_2$ and gauge symmetry implies that each bi-doublet will occur twice in $V_{bi}$. In the above we have only written some of the terms of $V_{bi}$.  
  
\begin{acknowledgments}
This work would not be possible without the solidarity of my friends and family.
\end{acknowledgments}
\bibliography{ravibib2}
\end{document}